\documentclass[prd,aps,showpacs,epsf,floats,10pt]{revtex4}%
\usepackage{amssymb}
\usepackage{amsfonts}
\usepackage{amsmath}
\usepackage{graphicx}

\providecommand{\U}[1]{\protect\rule{.1in}{.1in}}

\begin{document}
\title{ON RELATIVISTIC QUANTUM INFORMATION PROPERTIES OF ENTANGLED WAVE VECTORS OF
MASSIVE FERMIONS}
\author{Carlo Cafaro$^{1}$, Salvatore Capozziello$^{2\text{,}3}$, Stefano Mancini$^{1\text{,
}4}$}
\affiliation{$^{1}$School of Science and Technology, Physics Division, University of
Camerino, I-62032 Camerino, Italy}
\affiliation{$^{2}$Dipartimento di Scienze Fisiche, Universit\`{a} di Napoli
\textquotedblleft Federico II\textquotedblright\ and $^3$INFN Sez. di Napoli,
Compl. Univ. di Monte S. Angelo, Edificio G, Via Cinthia, 80126, Napoli, Italy}
\affiliation{$^{4}$INFN, Sezione di Perugia, I-06123 Perugia, Italy}

\begin{abstract}
We study special\textbf{ }relativistic effects on the entanglement between
either spins or momenta of composite quantum systems of two spin-$\frac{1}{2}$
massive particles, either indistinguishable or distinguishable, in inertial
reference frames in relative motion. For the case of indistinguishable
particles, we consider a balanced scenario where the momenta of the pair are
well-defined but not maximally entangled in the rest frame while the spins of
the pair are described by a one-parameter ($\eta$) family of entangled
bipartite states. For the case of distinguishable particles, we consider an
unbalanced scenario where the momenta of the pair are well-defined and
maximally entangled in the rest frame while the spins of the pair are
described by a one-parameter ($\xi$) family of non-maximally entangled
bipartite states. In both cases, we show that neither the spin-spin ($ss$) nor
the momentum-momentum ($mm$) entanglements quantified by means of Wootters'
concurrence are Lorentz invariant quantities: the total amount of entanglement
regarded as the sum of these entanglements is not the same in different
inertial moving frames. In particular, for any value of the entangling
parameters, both $ss$ and $mm$-entanglements are attenuated by Lorentz
transformations and their parametric rates of change with respect to the
entanglements observed in a rest frame have the same monotonic behavior.
However, for indistinguishable (distinguishable) particles, the change in
entanglement for the momenta is (is not) the same as the change in
entanglement for spins. As a consequence, in both cases, no entanglement
compensation between spin and momentum degrees of freedom occurs.

\end{abstract}

\maketitle

\section{Introduction}

It is known in relativistic thermodynamics that the concept of temperature is
observer-dependent because radiation that is perfectly black-body in a given
inertial reference frame is not thermal when viewed from a different moving
reference frame \cite{peebles, matsas}. Therefore, since probability
distributions can depend on the frames, many other information theoretic
quantities such as the Shannon entropy can exhibit such dependence. Quantum
information theory usually involves only a nonrelativistic quantum mechanics.
However, relativistic effects are of great importance also in quantum
information theory \cite{peres-terno}. From a practical point of view,
relativistic quantum information is important since it may provide a useful
theoretical platform for several possible applications such as
quantum-enhanced global positioning \cite{giovanetti}, quantum clock
synchronization \cite{jozsa} and quantum teleportation \cite{bennett}. From a
foundational viewpoint, relativistic extensions of quantum information theory
come from quantum cosmology \cite{peres-terno}. Specifically, quantum field
theory in curved spacetime (black hole physics, in particular) present
challenges that everybody who upholds the principle that "information is
physical" \cite{landauer} should respond to. A first consequence of relativity
on quantum theory is the existence of an upper bound, the velocity of light,
on the speed of propagation of physical effects. A more important consequence
of relativity is that there is a hierarchy of dynamical variables
\cite{peres-terno2}: \textit{primary variables} have relativistic
transformation laws that depend only on the Lorentz transformation matrix that
acts on the spacetime coordinates. For example, momentum components are
primary variables. On the other hand, \textit{secondary variables} such as
spin and polarization have transformation laws that depend not only on the
Lorentz transformation matrix, but also on the momentum of the particle. As a
consequence, the reduced density matrix for secondary variables, which may be
well defined in any coordinate system, has no transformation law relating its
components in different Lorentz frames. In relativistic quantum information
theory, the notion "spin state of a particle" is meaningless if we don't
specify its complete state, including the momentum variables
\cite{peres-scudo-terno}. It is possible to formally define spin in any
Lorentz frame, but there is no relationship between the observable expectation
values in different Lorentz frames. Stated otherwise, the answers to such
questions, asked in different Lorentz frames, are not related by any
transformation group. Under a Lorentz boost, the spin undergoes a Wigner
rotation whose direction and magnitude depend on the momentum of the particle.
Even if the initial state is a direct product of a function of momentum and a
function of spin, the transformed state is not a direct product. Spin and
momentum appear to be entangled. This implies that the spin entropy is not a
Lorentz scalar and has no invariant meaning in special relativity
\cite{peres-scudo-terno}.

Entanglement is one of the key features of quantum mechanics and a deep
understanding of its properties and implementations is essential not only to
fundamentally advance our understanding of how Nature works, but also to
design more powerful technologies. Entanglement is a property unique to
quantum systems. Two systems (microscopic particles or even macroscopic
bodies) are said to be quantum entangled if they are described by a joint wave
function that cannot be written as a product of wave functions of each of the
subsystems (or, for mixed states, if a density matrix cannot be written as a
weighted sum of product density matrices). The subsystems can be said not to
have a state of their own, even though they may be arbitrarily far apart. The
entanglement produces correlations between the subsystems that go beyond what
is classically possible. From a relativistic point of view, entanglement is an
observer-dependent concept \cite{ortiz}. In particular since Lorentz boosts
entangle the spin and momentum degrees of freedom, entanglement may be
transferred between them. This may be true for single particles
\cite{peres-scudo-terno} (where entanglement is being considered between
degrees of freedom belonging to the same particle), and for pairs
\cite{alsing, adami, ahn, soo}, where the Lorentz boost affects the
entanglement between spins. Within such quantum relativistic framework, qubits
are realized as discrete degrees of freedom of particles: a qubit can be
either a spin of a massive particle or a polarization of a photon. Once the
relativistic transformations of the states of massive particles and photons
are given, it can be deduced what happens to them when described by observers
in relative motion.

{In \cite{adami}, it was stated that the increase in spin entanglement comes
at the expense of a loss of momentum entanglement, since the entanglement
between all degrees of freedom (spin and momentum) is constant under Lorentz
transformations. The misinterpretation of this statement has lead to several
misleading remarks in the literature as pointed out in \cite{suda}. Actually
in \cite{suda} it was found that the change in entanglement for the momenta is
indeed the same as the change in entanglement for spins and no entanglement
transfer occurs. However, there, it was considered a pair of spin-}$\frac
{1}{2}$ {indistinguishable particles with maximally entangled momenta and
spins in the rest frame. Hence, the possibility of entanglement transfer
between different degrees of freedom is jeopardized by this assumption (it
automatically implies that moving from the rest frame, both }$ss$ and
$mm$-entanglements will diminish).{ This led us to consider more general
scenarios as explained in the following paragraph.}

In this article, we investigate special\textbf{ }relativistic effects on the
entanglement between either spins or momenta of composite quantum system of
two spin-$\frac{1}{2}$ massive particles, either indistinguishable or
distinguishable, in inertial reference frames in relative motion.
Specifically, for the case of indistinguishable particles, we consider a
balanced scenario where the momenta of the pair are well-defined but not
maximally entangled in the rest frame while the spins of the pair are
described by a one-parameter ($\eta$) family of non-maximally entangled
bipartite states. In particular, when $\eta=0$ (symmetry in the spin-wave
function and antisymmetry in the momentum-wave function) or $\eta=1$
(antisymmetry in the spin-wave function and symmetry in the momentum-wave
function), we recover the main result appeared in {\cite{suda}. }For the case
of distinguishable particles, we take into consideration an unbalanced
scenario where the momenta of the pair are well-defined and maximally
entangled in the rest frame while the spins of the pair are described by a
one-parameter ($\xi$) family of non-maximally entangled bipartite states.
{Furthermore, }in both cases, we show that neither the spin-spin ($ss$) nor
the momentum-momentum ($mm$) entanglements quantified by means of Wootters'
concurrence are Lorentz invariant quantities: the total amount of entanglement
regarded as the sum of these entanglements is not the same in different
inertial moving frames. In particular, for any value of the entangling
parameters, both $ss$ and $mm$-entanglements are attenuated by Lorentz
transformations and their parametric rates of change with respect to the
entanglements observed in a rest frame have the same monotonic behavior.
However, for indistinguishable (distinguishable) particles, the change in
entanglement for the momenta is (is not) the same as the change in
entanglement for spins. We conclude that in both cases no entanglement
compensation between spin and momentum degrees of freedom occurs.

The layout of the article is as follows. In Section II, we present preliminary
material on relativistic Lorentz transformations for quantum wave-vectors. In
Section III, we define the reduced density matrices of the considered
composite quantum systems of two spin-$\frac{1}{2}$ particles (either
indistinguishable or distinguishable) in a rest frame and analyze the action
of relativistic Lorentz boosts on them. In Section IV, we present the
relativistic effects on the entanglement of a composite system of two
indistinguishable particles (two electrons with wave-vector $\left\vert
\Psi_{ee}\right\rangle $) with a balanced (but not maximal) amount of
entanglement in the rest frame between momentum and spin. In Section V, we
analyze the relativistic effects on the entanglement of a composite system of
two distinguishable particles (an electron and a muon with wave-vector
$\left\vert \Psi_{e\mu}\right\rangle $) in which we consider an unbalanced
scenario. Our concluding remarks appear in Section VI.

\section{Lorentz transformations for quantum wave-vectors}

In \cite{wigner}, Wigner showed that quantum states of relativistic particles
are given by unitary irreducible representations of the Poincar\'{e} group,
the group of translations and Lorentz transformations (boosts) in the
Minkowski space. From \cite{wigner} it is also clear that finite dimensional
representations of Lorentz boosts are non-unitary. However, special relativity
requires that the physics of quantum states of relativistic particles should
not depend on the arbitrary inertial reference frame from which the states are
observed. Thus, we should expect the states to transform unitarily from one
inertial reference frame to another. The solution to this paradox resides in
the fact that in relativistic quantum mechanics, the creation and annihilation
operators, as well as the associated mode functions for the quantum field that
creates a given state, transform under Lorentz transformations by local
unitary spin-$j$ representations of the three-dimensional rotation group. The
key ingredient of such relativistic transformations is the Wigner rotation, a
rotation in the rest frame of the particle that leaves the rest momentum
invariant, which restores unitarity in the transformations between
relativistic single and multi-particle quantum states. Given the explicit
expression of the Wigner rotation, the transformation properties of entangled
quantum states observed from two inertial reference frames moving with
constant relative velocity can be described.

An arbitrary transformation $L\left(  \Lambda\text{, }a\right)  $ of the
Poincar\'{e} group (or, inhomogeneous Lorentz group), where $\Lambda$ denotes
a Lorentz transformation and $a$ a constant vector, relates the coordinates
$x^{\mu}$ and $x^{\prime\mu}$ of two inertial reference frames $S$ and
$S^{\prime}$ \cite{weinberg},%
\begin{equation}
x^{\mu}\longrightarrow x^{\prime\mu}\equiv L\left(  \Lambda\text{, }a\right)
x^{\mu}\overset{\text{def}}{=}\Lambda_{\nu}^{\mu}x^{\nu}+a^{\mu}\text{.}%
\end{equation}
The transformation $L\left(  \Lambda\text{, }a\right)  $ satisfy the
composition law,%
\begin{equation}
L\left(  \bar{\Lambda}\text{, }\bar{a}\right)  L\left(  \Lambda\text{,
}a\right)  =L\left(  \bar{\Lambda}\Lambda\text{, }\bar{\Lambda}a+\bar
{a}\right)  \text{,} \label{l}%
\end{equation}
where the bar in (\ref{l}) is used just to distinguish one Lorentz
transformation from the other. Moreover, $L\left(  \Lambda\text{, }a\right)  $
induces a linear unitary transformation $U\left(  \Lambda\text{, }a\right)  $
on quantum mechanical wave-vectors $\left\vert \Psi\right\rangle $,%
\begin{equation}
\left\vert \Psi\right\rangle \longrightarrow\left\vert \Psi^{\prime
}\right\rangle \equiv U\left(  \Lambda\text{, }a\right)  \left\vert
\Psi\right\rangle \text{,} \label{t}%
\end{equation}
where $U\left(  \Lambda\text{, }a\right)  $ satisfies the following
composition rule,%
\begin{equation}
U\left(  \bar{\Lambda}\text{, }\bar{a}\right)  U\left(  \Lambda\text{,
}a\right)  =U\left(  \bar{\Lambda}\Lambda\text{, }\bar{\Lambda}a+\bar
{a}\right)  \text{.}%
\end{equation}
For the sake of clarity, let us focus on single particle states and for a more
general approach we refer to \cite{weinberg}. Assuming to consider only
transformations with $a^{\mu}=0$ in the homogeneous Lorentz group, $\left\vert
\Psi^{\prime}\right\rangle $ in (\ref{t}) for a single particle state reads,%
\begin{equation}
\left\vert \Psi\right\rangle \longrightarrow\left\vert \Psi^{\prime
}\right\rangle \equiv U\left(  \Lambda\right)  \left\vert \Psi\right\rangle
\overset{\text{def}}{=}\sqrt{\frac{\left(  \Lambda p\right)  ^{0}}{p^{0}}}%
\sum_{\sigma^{\prime}}\mathcal{D}_{\sigma^{\prime}\sigma}^{\left(  j\right)
}\left(  W\left(  \Lambda\text{, }p\right)  \right)  \left\vert \Psi_{\Lambda
p\text{, }\sigma^{\prime}}\right\rangle \text{,} \label{psi}%
\end{equation}
where $p$ and $\sigma$ labels the momentum and the spin (or helicity for
massless particles) degrees of freedom, respectively. The coefficients
$\mathcal{D}_{\sigma^{\prime}\sigma}^{\left(  j\right)  }\left(  W\left(
\Lambda\text{, }p\right)  \right)  $ provide a representation of the Wigner
rotation $W\left(  \Lambda\text{, }p\right)  $ belonging to the so-called
Wigner's \textit{little group} for angular-$j$ particles (for instance,
$j=\frac{1}{2}$ for spin-$\frac{1}{2}$ particles). The Wigner's little group
element $W\left(  \Lambda\text{, }p\right)  $ is defined as,%
\begin{equation}
W\left(  \Lambda\text{, }p\right)  \overset{\text{def}}{=}L^{-1}\left(
\Lambda p\right)  \Lambda L\left(  p\right)  \text{,}%
\end{equation}
where $p^{\mu}\overset{\text{def}}{=}\left(  \vec{p}\text{, }p^{0}\right)  $
with,%
\begin{equation}
\vec{p}=\frac{m\vec{v}}{\sqrt{1-\left(  \frac{v}{c}\right)  ^{2}}}\text{ and
}p^{0}=\frac{E}{c}=\frac{mc}{\sqrt{1-\left(  \frac{v}{c}\right)  ^{2}}%
}\text{,}%
\end{equation}
and where $\left(  \Lambda p\right)  ^{\mu}\overset{\text{def}}{=}\left(
\vec{p}_{\Lambda}\text{, }\left(  \Lambda p\right)  ^{0}\right)  $ with
$\mu=0$, $1$, $2$, $3$. The quantity $L\left(  p\right)  $ is the standard
Lorentz transformation such that $p^{\mu}=L_{\nu}^{\mu}k^{\nu}$ where $k^{\nu
}=\left(  0\text{, }0\text{, }0\text{, }m\right)  $ is the four-momentum in
the rest frame of the particle being considered. Stated otherwise, $L\left(
p\right)  $ takes a massive particle from rest to a $4$-momentum $p$.

For massive ($m>0$) spin-$\frac{1}{2}$ particle states it turns out that
$\mathcal{D}^{\left(  1/2\right)  }\left(  W\left(  \Lambda\text{, }p\right)
\right)  $ in (\ref{psi}) reads \cite{halpern},%
\begin{equation}
\mathcal{D}^{\left(  1/2\right)  }\left(  W\left(  \Lambda\text{, }p\right)
\right)  =\mathcal{D}^{\left(  -1/2\right)  }\left(  L\left(  \Lambda
p\right)  \right)  \mathcal{D}^{\left(  1/2\right)  }\left(  \Lambda\right)
\mathcal{D}^{\left(  1/2\right)  }\left(  L\left(  p\right)  \right)  \text{.}%
\end{equation}
Consider an arbitrary boost characterized by the velocity $\vec{v}=v\hat{e}$
with $\hat{e}$ the normal vector in the boost direction. Then, omitting
tedious technical details that can be found in \cite{ahn, halpern}, it turns
out that the explicit expression of $\mathcal{D}^{\left(  1/2\right)  }\left(
W\left(  \Lambda\text{, }p\right)  \right)  $ reads,%
\begin{equation}
\mathcal{D}^{\left(  1/2\right)  }\left(  W\left(  \Lambda\text{, }p\right)
\right)  =\cos\frac{\Omega_{\vec{p}}}{2}+i\left(  \vec{\sigma}\cdot\hat
{n}\right)  \sin\frac{\Omega_{\vec{p}}}{2}\text{,}%
\end{equation}
where,%
\begin{equation}
\cos\frac{\Omega_{\vec{p}}}{2}=\frac{\cosh\frac{\alpha}{2}\cosh\frac{\delta
}{2}+\sinh\frac{\alpha}{2}\sinh\frac{\delta}{2}\left(  \hat{e}\cdot\hat
{p}\right)  }{\left[  \frac{1}{2}+\frac{1}{2}\cosh\alpha\cosh\delta+\frac
{1}{2}\left(  \hat{e}\cdot\hat{p}\right)  \sinh\alpha\sinh\delta\right]
^{\frac{1}{2}}}\text{,} \label{A}%
\end{equation}
and,%
\begin{equation}
\hat{n}\sin\frac{\Omega_{\vec{p}}}{2}=\frac{\sinh\frac{\alpha}{2}\sinh
\frac{\delta}{2}\left(  \hat{e}\times\hat{p}\right)  }{\left[  \frac{1}%
{2}+\frac{1}{2}\cosh\alpha\cosh\delta+\frac{1}{2}\left(  \hat{e}\cdot\hat
{p}\right)  \sinh\alpha\sinh\delta\right]  ^{\frac{1}{2}}}\text{.} \label{B}%
\end{equation}
The quantities $\alpha$ and $\delta$ in (\ref{A}) and (\ref{B}) are such that,%
\begin{equation}
\cosh\alpha=\frac{1}{\sqrt{1-\beta^{2}}}\text{ and, }\cosh\delta=\frac{p^{0}%
}{m}\text{,}%
\end{equation}
respectively, where $\beta\overset{\text{def}}{=}\frac{v}{c}$ with $c$ the
speed of light. The non-relativistic limit is recovered for $\beta=0$. A more
detailed description of relativistic transformations of either massive or
massless arbitrary multi-particle quantum states can be found in \cite{alsing,
weinberg}.

\section{General framework: two-particles composite quantum systems}

In this Section, we study the relativistic action of Lorentz boosts on the
density operators of the composite quantum systems being considered in a rest
frame $S$.

We consider two spin-$\frac{1}{2}$ particles $A$ and $B$, either
indistinguishable or distinguishable, with positive mass. We focus on
two-particle wave-vectors that in a given inertial reference frame at rest $S$
may be written as,%
\begin{equation}
\left\vert \Psi\right\rangle _{S}\equiv\left\vert \Psi_{p_{A}\text{, }%
\sigma_{A}\text{; }p_{B}\text{, }\sigma_{B}\text{ }}\right\rangle =\left\vert
\vec{p}_{A}\text{, }\sigma_{A}\text{; }\vec{p}_{B}\text{, }\sigma_{B}\text{
}\right\rangle \overset{\text{def}}{=}\left\vert \vec{p}_{A}\text{, }\vec
{p}_{B}\right\rangle \otimes\left\vert \sigma_{A}\text{, }\sigma
_{B}\right\rangle \text{,}%
\end{equation}
where $p$ and $\sigma$ denote the momentum and spin degrees of freedom of each
particle. Following \cite{suda}, we assume that the momentum of particle $A$
is concentrated around $\vec{p}_{A}$ and that of particle $B$ around $\vec
{p}_{B}$. Thus, the state for the momenta of the composite system $A+B$ is a
product state that reads,%
\begin{equation}
\left\vert \vec{p}_{A}\text{, }\vec{p}_{B}\right\rangle =\left\vert \vec
{p}_{A}\right\rangle _{A}\otimes\left\vert \text{ }\vec{p}_{B}\right\rangle
_{B}\text{.}%
\end{equation}
Furthermore, $\left\vert \sigma_{A}\text{, }\sigma_{B}\right\rangle $
represents a state for the spins of the two particles. In a different
reference frame $S^{\prime}$ in relative motion with respect to the rest frame
$S$, the composite wave-vector $\left\vert \Psi\right\rangle _{S}$ becomes%
\begin{align}
\left\vert \Psi\right\rangle _{S}\overset{S\rightarrow S^{\prime}%
}{\longrightarrow}\left\vert \Psi\right\rangle _{S^{\prime}}  &  =\sum
_{\sigma_{A}^{\prime}\text{, }\sigma_{B}^{\prime}}\mathcal{D}_{\sigma
_{A}^{\prime}\sigma_{A}}^{\left(  1/2\right)  }\left(  W\left(  \Lambda\text{,
}p_{A}\right)  \right)  \mathcal{D}_{\sigma_{B}^{\prime}\sigma_{B}}^{\left(
1/2\right)  }\left(  W\left(  \Lambda\text{, }p_{B}\right)  \right)
\left\vert \Lambda\vec{p}_{A}\text{, }\sigma_{A}^{\prime}\text{; }\Lambda
\vec{p}_{B}\text{, }\sigma_{B}^{\prime}\text{ }\right\rangle \nonumber\\
& \nonumber\\
=  &  \left\vert \vec{p}_{A}\text{, }\vec{p}_{B}\right\rangle ^{\Lambda
}\otimes\mathcal{D}_{A}^{\left(  1/2\right)  }\left(  W\left(  \Lambda\text{,
}p_{A}\right)  \right)  \mathcal{D}_{B}^{\left(  1/2\right)  }\left(  W\left(
\Lambda\text{, }p_{B}\right)  \right)  \left\vert \sigma_{A}\text{, }%
\sigma_{B}\right\rangle \text{,}%
\end{align}
where $\Lambda$ is the Lorentz boost between the two reference frames. The
transformed state for the momenta of the two particles is given by,%
\begin{equation}
\left\vert \vec{p}_{A}\text{, }\vec{p}_{B}\right\rangle \overset{S\rightarrow
S^{\prime}}{\longrightarrow}\left\vert \vec{p}_{A}\text{, }\vec{p}%
_{B}\right\rangle ^{\Lambda}=\left(  \left\vert \vec{p}_{A}\right\rangle
_{A}\otimes\left\vert \text{ }\vec{p}_{B}\right\rangle _{B}\right)  ^{\Lambda
}=\left\vert \vec{p}_{A}\right\rangle _{A}^{\Lambda}\otimes\left\vert \text{
}\vec{p}_{B}\right\rangle _{B}^{\Lambda}\text{.}%
\end{equation}
Moreover, the transformed state for the spins reads,%
\begin{equation}
\left\vert \sigma_{A}\text{, }\sigma_{B}\right\rangle \overset{S\rightarrow
S^{\prime}}{\longrightarrow}\mathcal{D}_{A}^{\left(  1/2\right)  }\left(
W\left(  \Lambda\text{, }p_{A}\right)  \right)  \mathcal{D}_{B}^{\left(
1/2\right)  }\left(  W\left(  \Lambda\text{, }p_{B}\right)  \right)
\left\vert \sigma_{A}\text{, }\sigma_{B}\right\rangle \text{,}%
\end{equation}
where $W\left(  \Lambda\text{, }p\right)  $ is the Wigner rotation for the
Lorentz transformation $\Lambda$ and $p^{\mu}=\left(  \vec{p}\text{,
}m\right)  $ with $m>0$. In the case under investigation, $\mathcal{D}%
^{\left(  1/2\right)  }\left(  W\left(  \Lambda\text{, }p\right)  \right)  $
is a $2\times2$ unitary (rotation) matrix.

In the rest of the manuscript, we focus our attention on the quantum Lorentz
transformation properties of the following two-particle quantum mechanical
state vector $\left\vert \Psi\right\rangle $ that, when viewed from the
reference rest frame $S$, reads%
\begin{equation}
\left\vert \Psi\right\rangle _{S}\equiv\left\vert \Psi\right\rangle =\frac
{1}{\sqrt{2}}\left[  \left\vert \vec{p}_{A_{1}}\text{, }\vec{p}_{B_{1}%
}\right\rangle \otimes\left\vert 0\right\rangle _{S}+\left\vert \vec{p}%
_{A_{2}}\text{, }\vec{p}_{B_{2}}\right\rangle \otimes\left\vert 0^{\prime
}\right\rangle _{S}\right]  \text{,} \label{prova}%
\end{equation}
where the state vectors $\left\vert 0\right\rangle _{S}$ and $\left\vert
0^{\prime}\right\rangle _{S}$ (with $\left\vert 0\right\rangle _{S}%
\neq\left\vert 0^{\prime}\right\rangle _{S}$, in general) describe the state
of spins of the massive fermions.

We stress that one of the main consequences of the spin-statistics theorem
\cite{pauli} is that the wave-function (both spin and space parts) of a system
of identical (indistinguishable) half-integer spin particles changes sign when
two particles are swapped. \ This can be achieved in two ways: either the
space wave-function is symmetric and the spin wave-function is antisymmetric,
or the space wave-function antisymmetric and the spin wave-function symmetric.
Fermions are particles whose wave-function is antisymmetric under exchange.
Observe that no measurement can distinguish between identical particles so we
must always have this exchange symmetry in the wave-function of identical
particles. For distinguishable particles that exhibit different intrinsic
physical properties, no exchange symmetry needs to hold.

From (\ref{prova}) we imply that just two pairs of momentum values $\left(
\vec{p}_{A_{1}}\text{, }\vec{p}_{B_{1}}\right)  $ and $\left(  \vec{p}_{A_{2}%
}\text{, }\vec{p}_{B_{2}}\right)  $ are needed to fully characterize the
momentum degrees of freedom of the composite system under investigation. As
pointed out earlier, we also consider the working hypothesis that such momenta
are concentrated closely enough around their distinct values so that this
allows us to use orthogonal state vectors for distinguishable concentrations.
Within such approximated scenario, we can employ a single Wigner rotation for
each concentration \cite{suda}.

In a reference frame $S^{\prime}$ in relative motion with respect to the rest
frame $S$, the quantum Lorentz transformation of the wave-vector $\left\vert
\Psi\right\rangle $ reads%
\begin{equation}
\left\vert \Psi\right\rangle \overset{S\rightarrow S^{\prime}}{\longrightarrow
}\left\vert \Psi\right\rangle _{S^{\prime}}\equiv\left\vert \Psi^{\prime
}\right\rangle =\frac{1}{\sqrt{2}}\left[  \left\vert \vec{p}_{A_{1}}\text{,
}\vec{p}_{B_{1}}\right\rangle ^{\Lambda}\otimes\left\vert 1\right\rangle
_{S^{\prime}}+\left\vert \vec{p}_{A_{2}}\text{, }\vec{p}_{B_{2}}\right\rangle
^{\Lambda}\otimes\left\vert 2\right\rangle _{S^{\prime}}\right]  \text{,}
\label{d}%
\end{equation}
where the transformed momenta state vector reads,%
\begin{equation}
\left\vert \vec{p}_{A_{1}}\text{, }\vec{p}_{B_{1}}\right\rangle ^{\Lambda
}\overset{\text{def}}{=}\left\vert \Lambda\vec{p}_{A_{1}}\text{, }\Lambda
\vec{p}_{B_{1}}\right\rangle \text{,} \label{a}%
\end{equation}
and the transformed spin states $\left\vert 1\right\rangle _{S^{\prime}}$ and
$\left\vert 2\right\rangle _{S^{\prime}}$ are given by,%
\begin{equation}
\left\vert 1\right\rangle _{S^{\prime}}\equiv\left\vert 1\right\rangle
\overset{\text{def}}{=}\mathcal{D}_{A}^{\left(  1/2\right)  }\left(  W\left(
\Lambda\text{, }p_{A_{1}}\right)  \right)  \mathcal{D}_{B}^{\left(
1/2\right)  }\left(  W\left(  \Lambda\text{, }p_{B_{1}}\right)  \right)
\left\vert 0\right\rangle _{S}=\mathcal{D}_{A}\left(  \text{ }p_{A_{1}%
}\right)  \mathcal{D}_{B}\left(  p_{B_{1}}\right)  \left\vert 0\right\rangle
_{S}\text{,} \label{b}%
\end{equation}
and,%
\begin{equation}
\left\vert 2\right\rangle _{S^{\prime}}\equiv\left\vert 2\right\rangle
\overset{\text{def}}{=}\mathcal{D}_{A}^{\left(  1/2\right)  }\left(  W\left(
\Lambda\text{, }p_{A_{2}}\right)  \right)  \mathcal{D}_{B}^{\left(
1/2\right)  }\left(  W\left(  \Lambda\text{, }p_{B_{2}}\right)  \right)
\left\vert 0^{\prime}\right\rangle _{S}=\mathcal{D}_{A}\left(  \text{
}p_{A_{2}}\right)  \mathcal{D}_{B}\left(  \text{ }p_{B_{2}}\right)  \left\vert
0^{\prime}\right\rangle _{S}\text{,} \label{c}%
\end{equation}
respectively. Substituting (\ref{a}), (\ref{b})\ and (\ref{c}) into (\ref{d}),
the Lorentz-transformed density operator becomes%
\begin{equation}
\hat{\rho}\overset{\text{def}}{=}\left\vert \Psi\right\rangle \left\langle
\Psi\right\vert \overset{S\rightarrow S^{\prime}}{\longrightarrow}\hat{\rho
}^{\prime}\overset{\text{def}}{=}\left\vert \Psi^{\prime}\right\rangle
\left\langle \Psi^{\prime}\right\vert =U\left(  \Lambda\right)  \hat{\rho
}U^{\dagger}\left(  \Lambda\right)  \text{,}%
\end{equation}
that is,%
\begin{equation}
\hat{\rho}^{\prime}=\frac{1}{2}\left[
\begin{array}
[c]{c}%
\left\vert \Lambda\vec{p}_{A_{1}}\text{, }\Lambda\vec{p}_{B_{1}}\right\rangle
\left\langle \Lambda\vec{p}_{A_{1}}\text{, }\Lambda\vec{p}_{B_{1}}\right\vert
\otimes\left\vert 1\right\rangle \left\langle 1\right\vert +\left\vert
\Lambda\vec{p}_{A_{1}}\text{, }\Lambda\vec{p}_{B_{1}}\right\rangle
\left\langle \Lambda\vec{p}_{A_{2}}\text{, }\Lambda\vec{p}_{B_{2}}\right\vert
\otimes\left\vert 1\right\rangle \left\langle 2\right\vert +\\
\\
+\left\vert \Lambda\vec{p}_{A_{2}}\text{, }\Lambda\vec{p}_{B_{2}}\right\rangle
\left\langle \Lambda\vec{p}_{A_{1}}\text{, }\Lambda\vec{p}_{B_{1}}\right\vert
\otimes\left\vert 2\right\rangle \left\langle 1\right\vert +\left\vert
\Lambda\vec{p}_{A_{2}}\text{, }\Lambda\vec{p}_{B_{2}}\right\rangle
\left\langle \Lambda\vec{p}_{A_{2}}\text{, }\Lambda\vec{p}_{B_{2}}\right\vert
\otimes\left\vert 2\right\rangle \left\langle 2\right\vert
\end{array}
\right]  \text{,}%
\end{equation}
where $\left\vert 1\right\rangle \equiv\left\vert 1\right\rangle _{S^{\prime}%
}$ and $\left\vert 2\right\rangle \equiv\left\vert 2\right\rangle _{S^{\prime
}}$. The reduced spin density operator is obtained from $\hat{\rho}^{\prime}$
by tracing over the momentum degrees of freedom and it reads%
\begin{align}
\hat{\rho}_{\text{spin}}^{\prime}\overset{\text{def}}{=}\text{Tr}%
_{\text{momentum}}\left(  \hat{\rho}^{\prime}\right)   &  =\frac{1}{2}\left[
\begin{array}
[c]{c}%
\left\langle \Lambda\vec{p}_{A_{1}}\text{, }\Lambda\vec{p}_{B_{1}}|\Lambda
\vec{p}_{A_{1}}\text{, }\Lambda\vec{p}_{B_{1}}\right\rangle \left\vert
1\right\rangle \left\langle 1\right\vert +\left\langle \Lambda\vec{p}_{A_{1}%
}\text{, }\Lambda\vec{p}_{B_{1}}|\Lambda\vec{p}_{A_{2}}\text{, }\Lambda\vec
{p}_{B_{2}}\right\rangle \left\vert 1\right\rangle \left\langle 2\right\vert
+\\
\\
+\left\langle \Lambda\vec{p}_{A_{2}}\text{, }\Lambda\vec{p}_{B_{2}}%
|\Lambda\vec{p}_{A_{1}}\text{, }\Lambda\vec{p}_{B_{1}}\right\rangle \left\vert
2\right\rangle \left\langle 1\right\vert +\left\langle \Lambda\vec{p}_{A_{2}%
}\text{, }\Lambda\vec{p}_{B_{2}}|\Lambda\vec{p}_{A_{2}}\text{, }\Lambda\vec
{p}_{B_{2}}\right\rangle \left\vert 2\right\rangle \left\langle 2\right\vert
\end{array}
\right] \nonumber\\
& \nonumber\\
&  =\frac{1}{2}\left[  \left\vert 1\right\rangle \left\langle 1\right\vert
+\left\vert 2\right\rangle \left\langle 2\right\vert \right]  \text{,}
\label{t-spin}%
\end{align}
with $\left\vert 1\right\rangle \equiv\left\vert 1\right\rangle _{S^{\prime}}$
and $\left\vert 2\right\rangle \equiv\left\vert 2\right\rangle _{S^{\prime}}$.
Similarly, the reduced momentum density operator is obtained from $\hat{\rho
}^{\prime}$ by tracing over the spin degrees of freedom and it reads,%
\begin{equation}
\hat{\rho}_{\text{momentum}}^{\prime}\overset{\text{def}}{=}\text{Tr}%
_{\text{spin}}\left(  \hat{\rho}^{\prime}\right)  =\frac{1}{2}\left[
\begin{array}
[c]{c}%
\left\langle 1|1\right\rangle \left\vert \Lambda\vec{p}_{A_{1}}\text{,
}\Lambda\vec{p}_{B_{1}}\right\rangle \left\langle \Lambda\vec{p}_{A_{1}%
}\text{, }\Lambda\vec{p}_{B_{1}}\right\vert +\left\langle 2|1\right\rangle
\left\vert \Lambda\vec{p}_{A_{1}}\text{, }\Lambda\vec{p}_{B_{1}}\right\rangle
\left\langle \Lambda\vec{p}_{A_{2}}\text{, }\Lambda\vec{p}_{B_{2}}\right\vert
+\\
\\
+\left\langle 1|2\right\rangle \left\vert \Lambda\vec{p}_{A_{2}}\text{,
}\Lambda\vec{p}_{B_{2}}\right\rangle \left\langle \Lambda\vec{p}_{A_{1}%
}\text{, }\Lambda\vec{p}_{B_{1}}\right\vert +\left\langle 2|2\right\rangle
\left\vert \Lambda\vec{p}_{A_{2}}\text{, }\Lambda\vec{p}_{B_{2}}\right\rangle
\left\langle \Lambda\vec{p}_{A_{2}}\text{, }\Lambda\vec{p}_{B_{2}}\right\vert
\end{array}
\right]  \text{.} \label{t-momentum}%
\end{equation}
The reduced density operators $\hat{\rho}_{\text{spin}}^{\prime}$ and
$\hat{\rho}_{\text{momentum}}^{\prime}$ in the moving reference frame
$S^{\prime}$ are the analogues of $\hat{\rho}_{\text{spin}}$ and $\hat{\rho
}_{\text{momentum}}$ in the rest reference frame $S$, respectively.

\section{Indistinguishable particles}

In this Section, we study the relativistic effects on the entanglement of a
composite system of two spin-$\frac{1}{2}$ indistinguishable massive particles
(two electrons in the state $\left\vert \Psi_{ee}\right\rangle $ in the rest
frame $S$) with a balanced (but not maximal) amount of entanglement in the
rest frame between momentum and spin.
\begin{figure}[t]
\begin{center}
\includegraphics{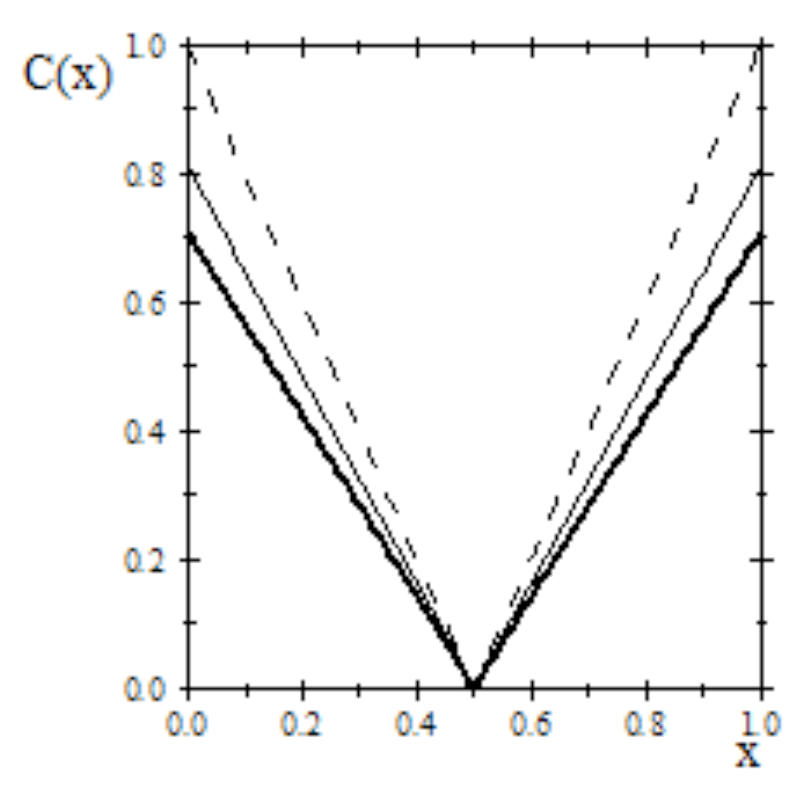}%
\caption{$\emph{C}_{\text{spin}}^{\left(  S^{\prime}\right)  }\left(
\eta\right)  \equiv\emph{C}_{\text{momentum}}^{\left(  S^{\prime}\right)
}\left(  \eta\right)  $ vs. $\eta$, $x\equiv\eta:$ $\varphi=0$ (dash),
$\varphi=\frac{\pi}{10}$ (thin solid), $\varphi=\frac{\pi}{8}$ (thick solid).}%
\end{center}
\end{figure}
We assume that the wave-vector $\left\vert \Psi_{S}\right\rangle $ in the rest
frame $S$ is given by,%
\begin{equation}
\left\vert \Psi_{ee}\right\rangle =\left\vert \Psi_{S}\right\rangle
\overset{\text{def}}{=}\frac{1}{\sqrt{2}}\left\vert \vec{p}_{A_{1}}\text{,
}\vec{p}_{B_{1}}\right\rangle \left[  \sqrt{\eta}\left\vert \phi
_{-}\right\rangle +\sqrt{1-\eta}\left\vert \phi_{+}\right\rangle \right]
+\frac{1}{\sqrt{2}}\left\vert \vec{p}_{A_{2}}\text{, }\vec{p}_{B_{2}%
}\right\rangle \left[  \sqrt{\eta}\left\vert \phi_{-}\right\rangle
-\sqrt{1-\eta}\left\vert \phi_{+}\right\rangle \right]  \text{,}\label{as}%
\end{equation}
where $\left\vert \phi_{\pm}\right\rangle $ are the maximally entangled
Bell-states,%
\begin{equation}
\left\vert \phi_{\pm}\right\rangle \overset{\text{def}}{=}\frac{\left\vert
\uparrow\downarrow\right\rangle \pm\left\vert \downarrow\uparrow\right\rangle
}{\sqrt{2}}=\frac{1}{\sqrt{2}}\left(
\begin{array}
[c]{c}%
0\\
1\\
\pm1\\
0
\end{array}
\right)  \text{,}\label{69}%
\end{equation}
and $\eta$ is a \emph{real} parameter such that $0\leq\eta\leq1$. Furthermore,
we assume that%
\begin{equation}
\vec{p}_{A_{1}}=-\vec{p}_{A_{2}}\text{, }\vec{p}_{B_{1}}=-\vec{p}_{B_{2}%
}\text{ and, }\vec{p}_{A_{1}}=\vec{p}_{B_{2}}\text{,}%
\end{equation}
where $\left\vert \vec{p}_{A_{1}}\right\rangle $ and $\left\vert \vec
{p}_{A_{2}}\right\rangle $ are the eigenvectors of $\tilde{\sigma}%
_{z}^{\left(  A\right)  }$ with eigenvalues $+1$ and $-1$, respectively, while
$\left\vert \vec{p}_{B_{1}}\right\rangle $ and $\left\vert \vec{p}_{B_{2}%
}\right\rangle $ are the eigenvectors of $\tilde{\sigma}_{z}^{\left(
B\right)  }$ with eigenvalues $-1$ and $+1$, respectively. Thus, we have%
\begin{equation}
\left\vert \vec{p}_{A_{1}}\right\rangle =\left\vert 0\right\rangle _{A}%
\equiv\binom{1}{0}\text{, }\left\vert \vec{p}_{A_{2}}\right\rangle =\left\vert
1\right\rangle _{A}\equiv\binom{0}{1}\text{, }\left\vert \vec{p}_{B_{1}%
}\right\rangle =\left\vert 1\right\rangle _{B}\equiv\binom{0}{1}\text{ and,
}\left\vert \vec{p}_{B_{2}}\right\rangle =\left\vert 0\right\rangle _{B}%
\equiv\binom{1}{0}\text{. }\label{as1}%
\end{equation}
Observe that from (\ref{as}) and (\ref{as1}), when $\eta=0$ and $\eta=1$,
$\left\vert \Psi_{S}\right\rangle $ in (\ref{as}) reduces to%
\begin{equation}
\left\vert \Psi_{+}\right\rangle \overset{\text{def}}{=}\frac{1}{\sqrt{2}%
}\left\vert 10\right\rangle \left\vert \phi_{+}\right\rangle -\frac{1}%
{\sqrt{2}}\left\vert 01\right\rangle \left\vert \phi_{+}\right\rangle \text{
and, }\left\vert \Psi_{-}\right\rangle \overset{\text{def}}{=}\frac{1}%
{\sqrt{2}}\left\vert 10\right\rangle \left\vert \phi_{-}\right\rangle
+\frac{1}{\sqrt{2}}\left\vert 01\right\rangle \left\vert \phi_{-}\right\rangle
\text{,}\label{sopra}%
\end{equation}
respectively. Using (\ref{as}) and (\ref{as1}), it follows that the reduced
spin density operator in the rest frame $S$ becomes (for further details, see
Appendix A),%
\begin{equation}
\rho_{\text{spin}}^{\left(  S\right)  }=\frac{1}{4}\left[  I_{2\times
2}^{\left(  A\right)  }\otimes I_{2\times2}^{\left(  B\right)  }+\left(
1-2\eta\right)  \sigma_{x}^{\left(  A\right)  }\otimes\sigma_{x}^{\left(
B\right)  }+\left(  1-2\eta\right)  \sigma_{y}^{\left(  A\right)  }%
\otimes\sigma_{y}^{\left(  B\right)  }-\sigma_{z}^{\left(  A\right)  }%
\otimes\sigma_{z}^{\left(  B\right)  }\right]  \text{,}%
\end{equation}
that is,%
\begin{equation}
\rho_{\text{spin}}^{\left(  S\right)  }\overset{\text{def}}{=}\text{Tr}%
_{\text{momentum}}\left(  \left\vert \Psi\right\rangle \left\langle
\Psi\right\vert \right)  =\frac{1}{2}\left(
\begin{array}
[c]{cccc}%
0 & 0 & 0 & 0\\
0 & 1 & 1-2\eta & 0\\
0 & 1-2\eta & 1 & 0\\
0 & 0 & 0 & 0
\end{array}
\right)  \text{.}%
\end{equation}
We quantify the entanglement between the spin degrees of freedom of particle
$A$ and $B$ by means of Wootter's concurrence \cite{wootters},%
\begin{equation}
\emph{C}\left(  \rho_{\text{spin}}^{\left(  S\right)  }\right)  \overset
{\text{def}}{=}\max\left\{  0\text{, }\lambda_{1}-\lambda_{2}-\lambda
_{3}-\lambda_{4}\right\}  \text{.}\label{34}%
\end{equation}
Denoting $\rho_{\text{spin}}^{\left(  S\right)  }\equiv\rho_{\text{spin}}$,
the non-negative \emph{real} numbers $\left\{  \lambda_{1}\text{, }\lambda
_{2}\text{, }\lambda_{3}\text{, }\lambda_{4}\right\}  $ denote the square
roots of the eigenvalues of the non-Hermitian matrix $\rho_{\text{spin}}%
\tilde{\rho}_{\text{spin}}$ where $\tilde{\rho}_{\text{spin}}$ is the
"time-reversed" matrix defined as,%
\begin{equation}
\tilde{\rho}_{\text{spin}}\overset{\text{def}}{=}\left(  \sigma_{y}^{\left(
A\right)  }\otimes\sigma_{y}^{\left(  B\right)  }\right)  \rho_{\text{spin}%
}^{\ast}\left(  \sigma_{y}^{\left(  A\right)  }\otimes\sigma_{y}^{\left(
B\right)  }\right)  \text{,}\label{35}%
\end{equation}
and $\rho_{\text{spin}}^{\ast}$ is the complex conjugate of $\rho
_{\text{spin}}$. It turns out that the concurrence $\emph{C}\left(
\rho_{\text{spin}}^{\left(  S\right)  }\right)  \equiv\mathcal{C}%
_{\text{spin}}^{\left(  S\right)  }\left(  \eta\text{, }\varphi\right)  $
reads,%
\begin{equation}
\mathcal{C}_{\text{spin}}^{\left(  S\right)  }\left(  \eta\text{, }%
\varphi\right)  =\left\vert \left(  1-2\eta\right)  \right\vert \text{.}%
\label{nota1}%
\end{equation}
Following the same line of reasoning, it can also be shown that the reduced
momentum density operator is given by,%
\begin{equation}
\rho_{\text{momentum}}^{\left(  S\right)  }=\frac{1}{4}\left[  I_{2\times
2}^{\left(  A\right)  }\otimes I_{2\times2}^{\left(  B\right)  }+\left(
2\eta-1\right)  \tilde{\sigma}_{x}^{\left(  A\right)  }\otimes\tilde{\sigma
}_{x}^{\left(  B\right)  }+\left(  2\eta-1\right)  \tilde{\sigma}_{y}^{\left(
A\right)  }\otimes\tilde{\sigma}_{y}^{\left(  B\right)  }-\tilde{\sigma}%
_{z}^{\left(  A\right)  }\otimes\tilde{\sigma}_{z}^{\left(  B\right)
}\right]  \text{,}%
\end{equation}
that is,%
\begin{equation}
\rho_{\text{momentum}}^{\left(  S\right)  }=\text{Tr}_{\text{spin}}\left(
\left\vert \Psi\right\rangle \left\langle \Psi\right\vert \right)  =\frac
{1}{2}\left(
\begin{array}
[c]{cccc}%
0 & 0 & 0 & 0\\
0 & 1 & 2\eta-1 & 0\\
0 & 2\eta-1 & 1 & 0\\
0 & 0 & 0 & 0
\end{array}
\right)  \text{.}\label{iuq}%
\end{equation}
The concurrence of $\rho_{\text{momentum}}^{\left(  S\right)  }$ in
(\ref{iuq}) becomes,%
\begin{equation}
\mathcal{C}_{\text{momentum}}^{\left(  S\right)  }\left(  \eta\text{, }%
\varphi\right)  =\left\vert \left(  1-2\eta\right)  \right\vert \text{.}%
\label{nota2}%
\end{equation}
From (\ref{nota1}) and (\ref{nota2}), we conclude that we are considering a
balanced but not maximal initial scenario as far as the entanglement in spins
and momenta concern.

In a moving reference frame $S^{\prime}$ where $\vec{p}_{A_{1}}$, $\vec
{p}_{A_{2}}$, $\vec{p}_{B_{1}}$ and $\vec{p}_{B_{2}}$ are along the $\hat{x}%
$-axis and the quantum Lorentz transformations are along the $\hat{y}%
$-direction with velocity $\vec{v}$, it turns out that the Lorentz-transformed
spin density operator in (\ref{t-spin}) reads,%
\begin{equation}
\rho_{\text{spin}}^{\left(  S^{\prime}\right)  }=\frac{1}{4}\left[
I_{2\times2}^{\left(  A\right)  }\otimes I_{2\times2}^{\left(  B\right)
}+\left(  1-2\eta\right)  \cos2\varphi\sigma_{x}^{\left(  A\right)  }%
\otimes\sigma_{x}^{\left(  B\right)  }+\left(  1-2\eta\right)  \cos
2\varphi\sigma_{y}^{\left(  A\right)  }\otimes\sigma_{y}^{\left(  B\right)
}-\sigma_{z}^{\left(  A\right)  }\otimes\sigma_{z}^{\left(  B\right)
}\right]  \text{,}%
\end{equation}
that is,%
\begin{equation}
\rho_{\text{spin}}^{\left(  S^{\prime}\right)  }\left(  \eta\text{, }%
\varphi\right)  =\frac{1}{2}\left(
\begin{array}
[c]{cccc}%
0 & 0 & 0 & 0\\
0 & 1 & \left(  1-2\eta\right)  \cos2\varphi & 0\\
0 & \left(  1-2\eta\right)  \cos2\varphi & 1 & 0\\
0 & 0 & 0 & 0
\end{array}
\right)  \text{.}%
\end{equation}
The concurrence $\mathcal{C}_{\text{spin}}^{\left(  S^{\prime}\right)
}\left(  \eta\text{, }\varphi\right)  $ becomes,%
\begin{equation}
\mathcal{C}_{\text{spin}}^{\left(  S^{\prime}\right)  }\left(  \eta\text{,
}\varphi\right)  =\left\vert \left(  1-2\eta\right)  \cos2\varphi\right\vert
\text{.} \label{nota3}%
\end{equation}
Similarly, the Lorentz-transformed momentum density operator in
(\ref{t-momentum}) is given by,%
\begin{equation}
\rho_{\text{momentum}}^{\left(  S^{\prime}\right)  }=\frac{1}{4}\left[
I_{2\times2}^{\left(  A\right)  }\otimes I_{2\times2}^{\left(  B\right)
}+\left(  2\eta-1\right)  \cos2\varphi\tilde{\sigma}_{x}^{\left(  A\right)
}\otimes\tilde{\sigma}_{x}^{\left(  B\right)  }+\left(  2\eta-1\right)
\cos2\varphi\tilde{\sigma}_{y}^{\left(  A\right)  }\otimes\tilde{\sigma}%
_{y}^{\left(  B\right)  }-\tilde{\sigma}_{z}^{\left(  A\right)  }\otimes
\tilde{\sigma}_{z}^{\left(  B\right)  }\right]  \text{,}%
\end{equation}
that is,%
\begin{equation}
\rho_{\text{momentum}}^{\left(  S^{\prime}\right)  }\left(  \eta\text{,
}\varphi\right)  =\frac{1}{2}\left(
\begin{array}
[c]{cccc}%
0 & 0 & 0 & 0\\
0 & 1 & \left(  2\eta-1\right)  \cos2\varphi & 0\\
0 & \left(  2\eta-1\right)  \cos2\varphi & 1 & 0\\
0 & 0 & 0 & 0
\end{array}
\right)  \text{.}%
\end{equation}
The concurrence $\mathcal{C}_{\text{momentum}}^{\left(  S^{\prime}\right)
}\left(  \eta\text{, }\varphi\right)  $ is given by,%
\begin{equation}
\mathcal{C}_{\text{momentum}}^{\left(  S^{\prime}\right)  }\left(  \eta\text{,
}\varphi\right)  =\left\vert \left(  1-2\eta\right)  \cos2\varphi\right\vert
\text{.} \label{nota4}%
\end{equation}
Thus, from (\ref{nota3}) and (\ref{nota4}), we uncover that in the case of
identical spin-$\frac{1}{2}$ particles,%
\begin{equation}
\mathcal{C}_{\text{spin}}^{\left(  S^{\prime}\right)  }\left(  \eta\text{,
}\varphi\right)  =\mathcal{C}_{\text{momentum}}^{\left(  S^{\prime}\right)
}\left(  \eta\text{, }\varphi\right)  \text{.} \label{same1}%
\end{equation}
We stress that when $\eta=0$ (symmetry in the spin-wave function and
antisymmetry in the momentum-wave function; first Eq. in (\ref{sopra})) or
$\eta=1$ (antisymmetry in the spin-wave function and symmetry in the
momentum-wave function; second Eq. in (\ref{sopra})), we recover the main
result appeared in {\cite{suda}. Finally, we emphasize that }for
indistinguishable particles, the change in entanglement for the momenta is the
same as the change in entanglement for spins (see Eq. (\ref{same1}), Figure
$1$ and Figure $2$).
\begin{figure}[t]
\begin{center}
\includegraphics{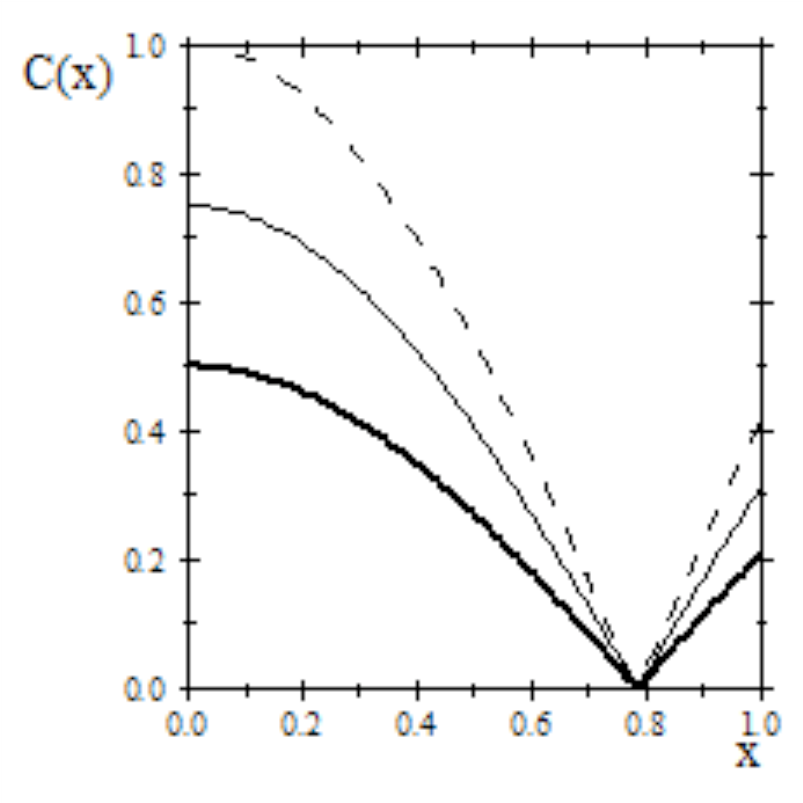}
\caption{$\emph{C}_{\text{spin}}^{\left(  S^{\prime}\right)  }\left(
\varphi\right)  \equiv\emph{C}_{\text{momentum}}^{\left(  S^{\prime}\right)
}\left(  \varphi\right)  $ vs. $\varphi$ for $x\equiv\varphi\in\left[
0\text{, }1\right]  :$ $\eta=0$ (dash), $\eta=\frac{1}{8}$ (thin solid),
$\eta=\frac{1}{4}$ (thick solid).}%
\end{center}
\end{figure}

\section{Distinguishable particles}

In this Section, we present a detailed study of relativistic effects on the
entanglement of distinguishable particles (an electron and a muon in the state
$\left\vert \Psi_{e\mu}\right\rangle $ in the rest frame $S$) where we
consider an unbalanced scenario where the momenta of the pair are well-defined
and maximally entangled in the rest frame while the spins of the pair are
described by a one-parameter ($\xi$) family of non-maximally entangled
bipartite states.

\subsection{Spin density operators in inertial reference frames in relative
motion}

We assume that the state vector $\left\vert 0\right\rangle _{S}=\left\vert
0^{\prime}\right\rangle _{S}\equiv$ $\left\vert 0\right\rangle $ in
(\ref{prova}) that describes the state of spins of the massive fermions is
given by,%
\begin{subequations}
\begin{equation}
\left\vert 0\right\rangle _{S}\equiv\left\vert 0\right\rangle =\left\vert
\psi_{\pm}\left(  \xi\right)  \right\rangle \overset{\text{def}}{=}\sqrt
{1-\xi}\left\vert \uparrow\downarrow\right\rangle \pm\sqrt{\xi}\left\vert
\downarrow\uparrow\right\rangle \text{,} \tag{47}\label{47}%
\end{equation}
with $0<\xi<1$ and where in the limiting case of $\xi=\frac{1}{2}$ we recover
standard maximally entangled pure Bell states. As a side remark, notice that
the matrix representation of the density operator to $\left\vert \psi_{\pm
}\left(  \xi\right)  \right\rangle $ reads%
\end{subequations}
\begin{equation}
\hat{\rho}_{\psi_{\pm}}\left(  \xi\right)  \overset{\text{def}}{=}\left\vert
\psi_{\pm}\left(  \xi\right)  \right\rangle \left\langle \psi_{\pm}\left(
\xi\right)  \right\vert =\left(
\begin{array}
[c]{cccc}%
0 & 0 & 0 & 0\\
0 & 1-\xi & \pm\sqrt{\xi\left(  1-\xi\right)  } & 0\\
0 & \pm\sqrt{\xi\left(  1-\xi\right)  } & \xi & 0\\
0 & 0 & 0 & 0
\end{array}
\right)  \text{.} \label{neo}%
\end{equation}
In order to make as clear as possible the effect of quantum Lorentz
transformation on quantum state vectors, it is convenient to re-express
$\hat{\rho}_{\psi_{\pm}}\left(  \xi\right)  $ in terms of Pauli matrices and
study the effect of Wigner rotations used to construct the Lorentz
transformations rotate Pauli's operators. It turns out that in terms of Pauli
matrices, $\hat{\rho}_{\psi_{\pm}}\left(  \xi\right)  $ in (\ref{neo}) may be
rewritten as (see Appendix A),%
\begin{equation}
\hat{\rho}_{\psi\pm}\left(  \xi\right)  =\frac{1}{4}\left[
\begin{array}
[c]{c}%
I_{2\times2}^{\left(  A\right)  }\otimes I_{2\times2}^{\left(  B\right)
}+\left(  1-2\xi\right)  \sigma_{z}^{\left(  A\right)  }\otimes I_{2\times
2}^{\left(  B\right)  }-\left(  1-2\xi\right)  I_{2\times2}^{\left(  A\right)
}\otimes\sigma_{z}^{\left(  B\right)  }\pm2\sqrt{\xi\left(  1-\xi\right)
}\sigma_{x}^{\left(  A\right)  }\otimes\sigma_{x}^{\left(  B\right)  }+\\
\\
\pm2\sqrt{\xi\left(  1-\xi\right)  }\sigma_{y}^{\left(  A\right)  }%
\otimes\sigma_{y}^{\left(  B\right)  }-\sigma_{z}^{\left(  A\right)  }%
\otimes\sigma_{z}^{\left(  B\right)  }%
\end{array}
\right]  \text{,} \label{ne}%
\end{equation}
where $\left(  \sigma_{x}^{\left(  A\right)  }\text{, }\sigma_{y}^{\left(
A\right)  }\text{, }\sigma_{z}^{\left(  A\right)  }\right)  $ and $\left(
\sigma_{x}^{\left(  B\right)  }\text{, }\sigma_{y}^{\left(  B\right)  }\text{,
}\sigma_{z}^{\left(  B\right)  }\right)  $ denote the Pauli matrices for the
spin of particle $A$ and $B$, respectively.

As a working hypothesis, we assume that,%
\begin{equation}
\vec{p}_{A_{1}}=-\vec{p}_{A_{2}}\text{ and }\vec{p}_{B_{1}}=-\vec{p}_{B_{2}%
}\text{ with }\vec{p}_{A_{1}}=-a\vec{p}_{B_{2}}\text{,} \label{set2}%
\end{equation}
where $\vec{p}_{A_{1}}$, $\vec{p}_{A_{2}}$, $\vec{p}_{B_{1}}$ and $\vec
{p}_{B_{2}}$ are along the $\hat{x}$-axis and the quantum Lorentz
transformations are along the $\hat{y}$-direction with velocity $\vec{v}$. The
constant \emph{real} coefficient $a$ must be such that $\varphi
_{\text{electron}}=\varphi_{\text{muon}}\equiv\varphi$. In the case under
investigation,%
\begin{equation}
\vec{p}_{A_{1}}=-a\vec{p}_{B_{2}}\Rightarrow p_{A_{1}}=ap_{B_{2}%
}\Longleftrightarrow\gamma\left(  v_{A_{1}}\right)  m_{e}v_{A_{1}}%
=a\gamma\left(  v_{B_{2}}\right)  m_{\mu}v_{B_{2}}\text{,}%
\end{equation}
that is (setting the speed of light equal to one, $c=1$),%
\begin{equation}
\frac{m_{e}v_{A_{1}}}{\sqrt{1-v_{A_{1}}^{2}}}=\frac{am_{\mu}v_{B_{2}}}%
{\sqrt{1-v_{B_{2}}^{2}}}\text{.}%
\end{equation}
After some algebraic manipulations, we obtain%
\begin{equation}
v_{A_{1}}=\frac{am_{\mu}v_{B_{2}}}{\sqrt{m_{e}^{2}+\left(  a^{2}m_{\mu}%
^{2}-m_{e}^{2}\right)  v_{B_{2}}^{2}}}=a\frac{m_{\mu}}{m_{e}}\frac{1}%
{\sqrt{1+\frac{a^{2}m_{\mu}^{2}-m_{e}^{2}}{m_{e}^{2}}v_{B_{2}}^{2}}}v_{B_{2}%
}\text{.} \label{ONE1}%
\end{equation}
In the case being considered, following the line of reasoning presented in
\cite{ahn}, the Wigner rotation angle $\varphi$ reads,%
\begin{equation}
\tan\varphi=\frac{\sinh\alpha\sinh\delta}{\cosh\alpha+\cosh\delta}\text{,}%
\end{equation}
where,%
\begin{equation}
\cosh\alpha=\frac{1}{\sqrt{1-\left(  \frac{v}{c}\right)  ^{2}}}=\frac{1}%
{\sqrt{1-v_{\text{boost}}^{2}}}\text{ and, }\cosh\delta=\frac{p^{0}}{m}%
=\frac{\gamma\left(  v_{\text{particle}}\right)  mc^{2}}{m}=\frac{1}%
{\sqrt{1-v_{\text{particle}}^{2}}}\text{.}%
\end{equation}
Therefore,%
\begin{equation}
\tan\varphi_{1}=\tan\varphi_{2}\Longleftrightarrow\frac{\sinh\alpha\sinh
\delta_{1}}{\cosh\alpha+\cosh\delta_{1}}=\frac{\sinh\alpha\sinh\delta_{2}%
}{\cosh\alpha+\cosh\delta_{2}}\Longleftrightarrow\frac{\sinh\delta_{1}}%
{\cosh\alpha+\cosh\delta_{1}}=\frac{\sinh\delta_{2}}{\cosh\alpha+\cosh
\delta_{2}}\text{,}%
\end{equation}
that is,%
\begin{equation}
\sinh\delta_{1}=\sinh\delta_{2}\Longleftrightarrow\frac{v_{\text{electron}}%
}{\sqrt{1-v_{\text{electron}}^{2}}}=\frac{v_{\text{muon}}}{\sqrt
{1-v_{\text{muon}}^{2}}}\Longleftrightarrow v_{\text{electron}}=v_{\text{muon}%
}\text{,} \label{TWO2}%
\end{equation}
where $v_{A_{1}}\equiv$ $v_{\text{electron}}$ and $v_{B_{2}}\equiv
v_{\text{muon}}$. Thus, from (\ref{ONE1}) and (\ref{TWO2}), internal
consistency requires%
\begin{equation}
v_{A_{1}}=a\frac{m_{\mu}}{m_{e}}\frac{1}{\sqrt{1+\frac{a^{2}m_{\mu}^{2}%
-m_{e}^{2}}{m_{e}^{2}}v_{B_{2}}^{2}}}v_{B_{2}}=v_{B_{2}}\text{,}%
\end{equation}
which is true provided that,%
\begin{equation}
a\frac{m_{\mu}}{m_{e}}\frac{1}{\sqrt{1+\frac{a^{2}m_{\mu}^{2}-m_{e}^{2}}%
{m_{e}^{2}}v_{B_{2}}^{2}}}=1\Longleftrightarrow a=\frac{m_{e}}{m_{\mu}%
}\text{.} \label{look}%
\end{equation}
In conclusion, we require%
\begin{equation}
\vec{p}_{A_{1}}=-\vec{p}_{A_{2}}\text{ and }\vec{p}_{B_{1}}=-\vec{p}_{B_{2}%
}\text{ with }\vec{p}_{A_{1}}=-\frac{m_{e}}{m_{\mu}}\vec{p}_{B_{2}}\text{.}%
\end{equation}
It finally turns out that the Lorentz transformations are defined in terms of
the following $\mathcal{D}\left(  p\right)  $ operators,%
\begin{align}
\mathcal{D}_{A}\left(  p_{A_{1}}\right)   &  =\cos\frac{\varphi}{2}-i\sin
\frac{\varphi}{2}\sigma_{z}^{\left(  A\right)  }\text{, }\mathcal{D}%
_{B}\left(  p_{B_{1}}\right)  =\cos\frac{\varphi}{2}+i\sin\frac{\varphi}%
{2}\sigma_{z}^{\left(  B\right)  }\text{,}\nonumber\\
& \nonumber\\
\text{ }\mathcal{D}_{A}\left(  p_{A_{2}}\right)   &  =\cos\frac{\varphi}%
{2}+i\sin\frac{\varphi}{2}\sigma_{z}^{\left(  A\right)  }\text{, }%
\mathcal{D}_{B}\left(  p_{B_{2}}\right)  =\cos\frac{\varphi}{2}-i\sin
\frac{\varphi}{2}\sigma_{z}^{\left(  B\right)  }\text{,} \label{D}%
\end{align}
where $\varphi$ is the angle that characterizes the Wigner rotation. From
(\ref{D}), it turns out that the Wigner rotations $W\left(  p_{A_{1}}\right)
$ and $W\left(  p_{B_{2}}\right)  $ are rotations by $\varphi$ around the
$\hat{z}$-axis and $W\left(  p_{A_{2}}\right)  $ and $W\left(  p_{B_{1}%
}\right)  $ are Wigner rotations by $-\varphi$ around the $\hat{z}$-axis.
Thus, substituting (\ref{D}) into (\ref{b}) and (\ref{c}) and using
(\ref{47}), the Lorentz transformed spin density operator $\hat{\rho
}_{\text{spin}}^{\prime}\equiv\hat{\rho}_{\psi\pm}^{\Lambda}\left(  \xi\text{,
}\varphi\right)  $ in (\ref{t-spin}) becomes (for further details, see
Appendix B),%
\begin{equation}
\hat{\rho}_{\psi\pm}^{\Lambda}\left(  \xi\text{, }\varphi\right)  =\frac{1}%
{4}\left[
\begin{array}
[c]{c}%
I_{2\times2}^{\left(  A\right)  }\otimes I_{2\times2}^{\left(  B\right)
}+\left(  1-2\xi\right)  \sigma_{z}^{\left(  A\right)  }\otimes I_{2\times
2}^{\left(  B\right)  }-\left(  1-2\xi\right)  I_{2\times2}^{\left(  A\right)
}\otimes\sigma_{z}^{\left(  B\right)  }\pm2\sqrt{\xi\left(  1-\xi\right)
}\cos2\varphi\sigma_{x}^{\left(  A\right)  }\otimes\sigma_{x}^{\left(
B\right)  }+\\
\\
\pm2\sqrt{\xi\left(  1-\xi\right)  }\cos2\varphi\sigma_{y}^{\left(  A\right)
}\otimes\sigma_{y}^{\left(  B\right)  }-\sigma_{z}^{\left(  A\right)  }%
\otimes\sigma_{z}^{\left(  B\right)  }\text{.}%
\end{array}
\right]  \text{,} \label{nue}%
\end{equation}
or, after some algebra,%
\begin{equation}
\hat{\rho}_{\psi\pm}^{\Lambda}\left(  \xi\text{, }\varphi\right)  =\hat{\rho
}_{\psi\pm}\left(  \xi\right)  \cos^{2}\varphi+\hat{\rho}_{\psi\mp}\left(
\xi\right)  \sin^{2}\varphi\text{,}%
\end{equation}
with $\hat{\rho}_{\psi\pm}\left(  \xi\right)  $ given in (\ref{ne}).

\subsection{Momentum density operators in inertial reference frames in
relative motion}%

\begin{figure}[t]
\begin{center}
\includegraphics{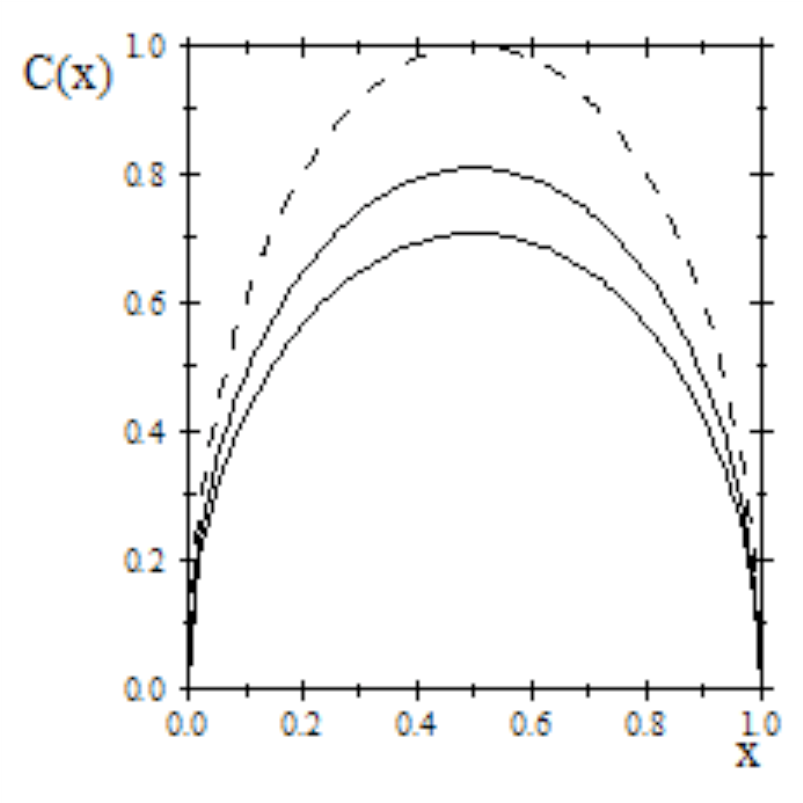}
\caption{$\emph{C}_{\text{spin}}^{\left(  S^{\prime}\right)  }\left(
\xi\right)  $ vs. $\xi$, $x\equiv\xi:$ $\varphi=0$ (dash), $\varphi=\frac{\pi
}{10}$ (thin solid), $\varphi=\frac{\pi}{8}$ (thick solid).}%
\end{center}
\end{figure}
In agreement with \cite{suda}, we regard the momentum states $\left\vert
\vec{p}_{A}\text{, }\vec{p}_{B}\right\rangle $ for the composite quantum
system under investigation as two-qubits momentum states. We use
$\mathbf{\vec{\Sigma}}^{\left(  A\right)  }=\left(  \tilde{\sigma}%
_{x}^{\left(  A\right)  }\text{, }\tilde{\sigma}_{y}^{\left(  A\right)
}\text{, }\tilde{\sigma}_{z}^{\left(  A\right)  }\text{ }\right)  $ and
$\mathbf{\vec{\Sigma}}^{\left(  B\right)  }=\left(  \tilde{\sigma}%
_{x}^{\left(  B\right)  }\text{, }\tilde{\sigma}_{y}^{\left(  B\right)
}\text{, }\tilde{\sigma}_{z}^{\left(  B\right)  }\text{ }\right)  $ to denote
the vectors of Pauli matrices used to describe the momentum qubit for
particles $A$ and $B$, respectively. Specifically, we assume that $\left\vert
\vec{p}_{A_{1}}\right\rangle $ and $\left\vert \vec{p}_{A_{2}}\right\rangle $
are the eigenvectors of $\tilde{\sigma}_{z}^{\left(  A\right)  }$ with
eigenvalues $+1$ and $-1$, respectively. Thus, we get%
\begin{equation}
\left\vert \vec{p}_{A_{1}}\right\rangle =\left\vert 0\right\rangle _{A}%
\equiv\binom{1}{0}\text{ and, }\left\vert \vec{p}_{A_{2}}\right\rangle
=\left\vert 1\right\rangle _{A}\equiv\binom{0}{1}\text{.}\label{pp1}%
\end{equation}
Similarly, we assume that $\left\vert \vec{p}_{B_{1}}\right\rangle $ and
$\left\vert \vec{p}_{B_{2}}\right\rangle $ are the eigenvectors of
$\tilde{\sigma}_{z}^{\left(  B\right)  }$ with eigenvalues $+1$ and $-1$,
respectively. Thus, we obtain%
\begin{equation}
\left\vert \vec{p}_{B_{1}}\right\rangle =\left\vert 0\right\rangle _{B}%
\equiv\binom{1}{0}\text{and, }\left\vert \vec{p}_{B_{2}}\right\rangle
=\left\vert 1\right\rangle _{B}\equiv\binom{0}{1}\text{.}\label{pp22}%
\end{equation}
Combining Eqs. (\ref{prova}), (\ref{47}), (\ref{pp1}) and (\ref{pp22}), the
quantum mechanical wave-vector in the rest frame $S$ becomes,%
\begin{equation}
\left\vert \Psi_{e\mu}\right\rangle =\left\vert \Psi_{S}\right\rangle
\overset{\text{def}}{=}\frac{1}{\sqrt{2}}\left\vert \vec{p}_{A_{1}}\text{,
}\vec{p}_{B_{1}}\right\rangle \left[  \sqrt{1-\xi}\left\vert \uparrow
\downarrow\right\rangle \pm\sqrt{\xi}\left\vert \downarrow\uparrow
\right\rangle \right]  +\frac{1}{\sqrt{2}}\left\vert \vec{p}_{A_{2}}\text{,
}\vec{p}_{B_{2}}\right\rangle \left[  \sqrt{1-\xi}\left\vert \uparrow
\downarrow\right\rangle \pm\sqrt{\xi}\left\vert \downarrow\uparrow
\right\rangle \right]  \text{.}\label{as0}%
\end{equation}
The Lorentz transformed quantum states corresponding to $\left\vert \vec
{p}_{A_{1}}\right\rangle $, $\left\vert \vec{p}_{A_{2}}\right\rangle $,
$\left\vert \vec{p}_{B_{1}}\right\rangle $ and $\left\vert \vec{p}_{B_{2}%
}\right\rangle $ are given by,%
\begin{align}
\left\vert \vec{p}_{A_{1}}\right\rangle \overset{S\rightarrow S^{\prime}%
}{\longrightarrow}\left\vert \vec{p}_{A_{1}}\right\rangle ^{\Lambda} &
=\left(  \cos\frac{\varphi}{2}-i\sin\frac{\varphi}{2}\tilde{\sigma}%
_{z}^{\left(  A\right)  }\right)  \left\vert 0\right\rangle _{A}\text{,
}\left\vert \vec{p}_{A_{2}}\right\rangle \overset{S\rightarrow S^{\prime}%
}{\longrightarrow}\left\vert \vec{p}_{A_{2}}\right\rangle ^{\Lambda}=\left(
\cos\frac{\varphi}{2}+i\sin\frac{\varphi}{2}\tilde{\sigma}_{z}^{\left(
A\right)  }\right)  \left\vert 1\right\rangle _{A}\text{,}\nonumber\\
& \nonumber\\
\left\vert \vec{p}_{B_{1}}\right\rangle \overset{S\rightarrow S^{\prime}%
}{\longrightarrow}\left\vert \vec{p}_{B_{1}}\right\rangle ^{\Lambda} &
=\left(  \cos\frac{\varphi}{2}+i\sin\frac{\varphi}{2}\tilde{\sigma}%
_{z}^{\left(  B\right)  }\right)  \left\vert 0\right\rangle _{B}\text{,
}\left\vert \vec{p}_{B_{2}}\right\rangle \overset{S\rightarrow S^{\prime}%
}{\longrightarrow}\left\vert \vec{p}_{B_{2}}\right\rangle ^{\Lambda}=\left(
\cos\frac{\varphi}{2}-i\sin\frac{\varphi}{2}\tilde{\sigma}_{z}^{\left(
B\right)  }\right)  \left\vert 1\right\rangle _{B}\text{.}\label{pp3}%
\end{align}
Upon computation of the inner products $\left\langle 1|2\right\rangle $ and
$\left\langle 2|1\right\rangle $ and using Eqs. (\ref{pp1}), (\ref{pp22}) and
(\ref{pp3}), it can be shown that $\hat{\rho}_{\text{momentum}}^{\prime}$ in
(\ref{t-momentum}) reads,%
\begin{equation}
\hat{\rho}_{\text{momentum}}^{\prime}\left(  \xi\text{, }\varphi\right)
=\frac{1}{4}\left[
\begin{array}
[c]{c}%
I_{2\times2}^{\left(  A\right)  }\otimes I_{2\times2}^{\left(  B\right)
}+\cos2\varphi\left(  \tilde{\sigma}_{x}^{\left(  A\right)  }\otimes
\tilde{\sigma}_{x}^{\left(  B\right)  }-\tilde{\sigma}_{y}^{\left(  A\right)
}\otimes\tilde{\sigma}_{y}^{\left(  B\right)  }\right)  +\tilde{\sigma}%
_{z}^{\left(  A\right)  }\otimes\tilde{\sigma}_{z}^{\left(  B\right)  }+\\
\\
+\left(  1-2\xi\right)  \sin2\varphi\left(  \tilde{\sigma}_{x}^{\left(
A\right)  }\otimes\tilde{\sigma}_{y}^{\left(  B\right)  }+\tilde{\sigma}%
_{y}^{\left(  A\right)  }\otimes\tilde{\sigma}_{x}^{\left(  B\right)
}\right)
\end{array}
\right]  \text{.}\label{derivation}%
\end{equation}
For further details on the derivation of (\ref{derivation}), we refer to
Appendix C.

\subsection{Entanglement in inertial reference frames in relative motion}

In what follows, we compute the concurrence of the reduced (mixed) density
matrices $\hat{\rho}_{\text{spin}}^{\prime}\left(  \xi\text{, }\varphi\right)
$ in (\ref{nue}) and $\hat{\rho}_{\text{momentum}}^{\prime}\left(  \xi\text{,
}\varphi\right)  $ in (\ref{derivation}).

\subsubsection{Concurrence of the Lorentz-transformed spin density matrix}%

\begin{figure}[t]
\begin{center}
\includegraphics{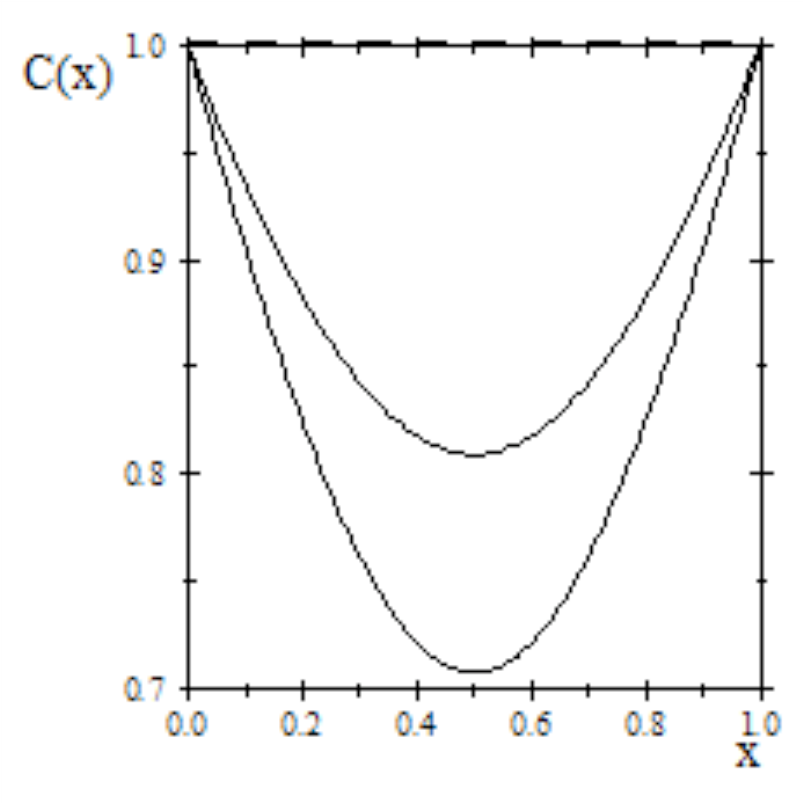}
\caption{$\emph{C}_{\text{momentum}}^{\left(  S^{\prime}\right)  }\left(
\xi\right)  $ vs. $\xi$, $x\equiv\xi:$ $\varphi=0$ (dash), $\varphi=\frac{\pi
}{10}$ (thin solid), $\varphi=\frac{\pi}{8}$ (thick solid).}%
\end{center}
\end{figure}
We quantify the entanglement between the spin degrees of freedom of particle
$A$ and $B$ by means of Wootter's concurrence \cite{wootters}. Using
(\ref{34}) and (\ref{35}), it follows that the concurrence of $\hat{\rho
}_{\psi_{\pm}}\left(  \xi\right)  $ in (\ref{neo}) reads%
\begin{equation}
\emph{C}_{\text{spin}}^{\left(  S\right)  }\left(  \xi\right)  =\sqrt
{4\xi(1-\xi)}\text{,}%
\end{equation}
where $\emph{C}_{\text{spin}}^{\left(  S\right)  }\left(  \xi\right)  $ is the
concurrence of the spin density matrix as viewed by an inertial observer
located in the reference rest frame $S$. In the moving reference frame
$S^{\prime}$, it turns out that the matrix representation of $\hat{\rho}%
_{\psi\pm}^{\Lambda}\left(  \xi\text{, }\varphi\right)  $ in (\ref{nue}) reads%
\begin{equation}
\hat{\rho}_{\psi\pm}^{\Lambda}\left(  \xi\text{, }\varphi\right)  =\left(
\begin{array}
[c]{cccc}%
0 & 0 & 0 & 0\\
0 & 1-\xi & \pm\sqrt{\xi\left(  1-\xi\right)  }\cos2\varphi & 0\\
0 & \pm\sqrt{\xi\left(  1-\xi\right)  }\cos2\varphi & \xi & 0\\
0 & 0 & 0 & 0
\end{array}
\right)  \text{,}\label{oggi}%
\end{equation}
and $\rho_{\text{spin}}\tilde{\rho}_{\text{spin}}$ with $\rho_{\text{spin}%
}=\hat{\rho}_{\psi\pm}^{\Lambda}\left(  \xi\text{, }\varphi\right)  $ in
(\ref{oggi}) becomes,%
\begin{equation}
\rho_{\text{spin}}\tilde{\rho}_{\text{spin}}=\left(
\begin{array}
[c]{cccc}%
0 & 0 & 0 & 0\\
0 & \xi\left(  1-\xi\right)  \left(  1+\cos^{2}2\varphi\right)   & \pm2\left(
1-\xi\right)  \sqrt{\xi\left(  1-\xi\right)  }\cos2\varphi & 0\\
0 & \pm2\xi\sqrt{\xi\left(  1-\xi\right)  }\cos2\varphi & \xi\left(
1-\xi\right)  \left(  1+\cos^{2}2\varphi\right)   & 0\\
0 & 0 & 0 & 0
\end{array}
\right)  \text{.}\label{roro}%
\end{equation}
The eigenvalues of $\rho_{\text{spin}}\tilde{\rho}_{\text{spin}}$ in
(\ref{roro}) are,%
\begin{equation}
\allowbreak\lambda_{1}=\xi\left(  1-\xi\right)  \left(  1+\cos2\varphi\right)
^{2}\text{, }\lambda_{2}=\xi\left(  1-\xi\right)  \left(  1-\cos
2\varphi\right)  ^{2}\text{, }\lambda_{3}=\lambda_{4}=0\text{.}%
\end{equation}
Finally, the concurrence of the Lorentz-transformed spin density matrix reads,%
\begin{equation}
\emph{C}_{\text{spin}}^{\left(  S^{\prime}\right)  }\left(  \xi\text{,
}\varphi\right)  =\sqrt{4\xi(1-\xi)\cos^{2}2\varphi}=\sqrt{4\xi(1-\xi
)}\left\vert \cos2\varphi\right\vert \text{.}\label{ONE}%
\end{equation}

\subsubsection{Concurrence of the Lorentz-transformed momentum density matrix}%

\begin{figure}[t]
\begin{center}
\includegraphics{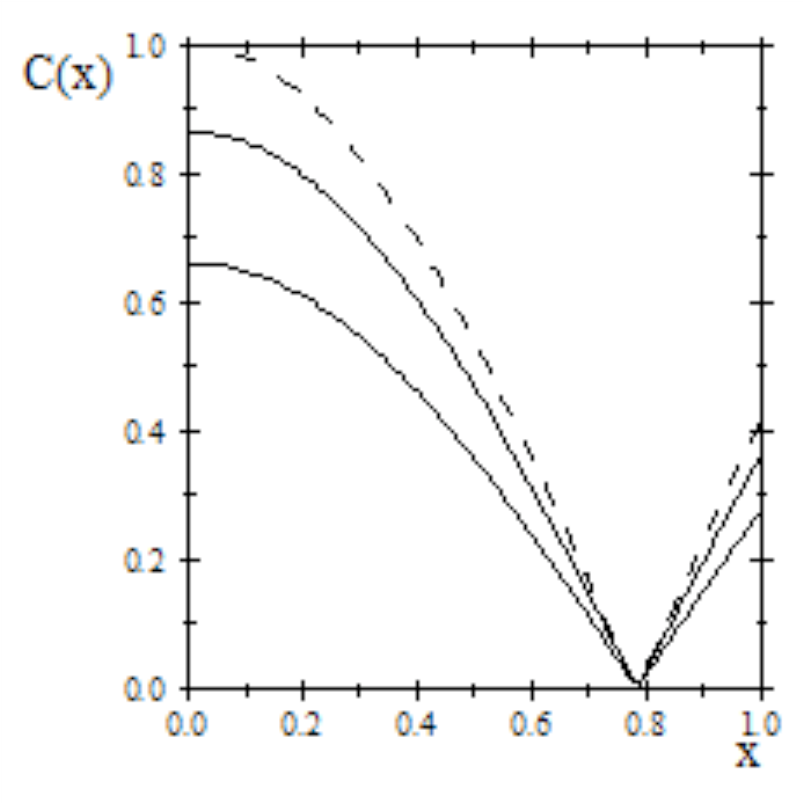}
\caption{$\emph{C}_{\text{spin}}^{\left(  S^{\prime}\right)  }\left(
\varphi\right)  $ vs. $\varphi$ for $x\equiv\varphi\in\left[  0\text{,
}1\right]  :$ $\xi=\frac{1}{2}$ (dash), $\xi=\frac{1}{4}$ (thin solid),
$\xi=\frac{1}{8}$ (thick solid).}%
\end{center}
\end{figure}
Following our analysis in the previous Subsection, we quantify the
entanglement between the momentum degrees of freedom of particle $A$ and $B$
by means of Wootter's concurrence $\emph{C}\left(  \rho_{\text{momentum}%
}\right)  $ where $\rho_{\text{momentum}}$ denotes the reduced density matrix
obtained by tracing over the spin degrees of freedom.

In the reference frame $S^{\prime}$, it turns out that the matrix
representation of $\hat{\rho}_{\text{momentum}}^{\prime}\left(  \xi\text{,
}\varphi\right)  $ in (\ref{derivation}) reads%
\begin{equation}
\hat{\rho}_{\text{momentum}}^{\prime}\left(  \xi\text{, }\varphi\right)
=\frac{1}{2}\left(
\begin{array}
[c]{cccc}%
1 & 0 & 0 & \cos2\varphi-i\left(  1-2\xi\right)  \sin2\varphi\\
0 & 0 & 0 & 0\\
0 & 0 & 0 & 0\\
\cos2\varphi+i\left(  1-2\xi\right)  \sin2\varphi & 0 & 0 & 1
\end{array}
\right)  \text{,}%
\end{equation}
and $\rho_{\text{momentum}}\tilde{\rho}_{\text{momentum }}$ with
$\rho_{\text{momentum}}=\hat{\rho}_{\text{momentum}}^{\prime}\left(
\xi\text{, }\varphi\right)  $ becomes,
\begin{equation}
\rho_{\text{momentum}}\tilde{\rho}_{\text{momentum }}=\frac{1}{2}\left(
\begin{array}
[c]{cccc}%
\xi\left(  \xi-1\right)  \left(  1-\cos2\varphi\right)  +1 & 0 & 0 &
\cos2\varphi-i\left(  1-2\xi\right)  \sin2\varphi\\
0 & 0 & 0 & 0\\
0 & 0 & 0 & 0\\
\cos2\varphi+i\left(  1-2\xi\right)  \sin2\varphi & 0 & 0 & \xi\left(
\xi-1\right)  \left(  1-\cos2\varphi\right)  +1
\end{array}
\right)  \text{.}\label{roro1}%
\end{equation}
It turns out that the eigenvalues of $\rho_{\text{momentum}}\tilde{\rho
}_{\text{momentum }}$ in (\ref{roro1}) are given by,%
\begin{equation}
\lambda_{1}=\left[  \frac{1}{2}\left(  1+\sqrt{1-4\xi\left(  1-\xi\right)
\sin^{2}2\varphi}\right)  \right]  ^{2}\text{, }\lambda_{2}=\left[  \frac
{1}{2}\left(  1-\sqrt{1-4\xi\left(  1-\xi\right)  \sin^{2}2\varphi}\right)
\right]  ^{2}\text{, }\lambda_{3}=\lambda_{4}=0\text{.}%
\end{equation}
Finally, the concurrence of the Lorentz-transformed momentum density matrix
reads,%
\begin{equation}
\emph{C}_{\text{momentum}}^{\left(  S^{\prime}\right)  }\left(  \xi\text{,
}\varphi\right)  =\sqrt{1-4\xi(1-\xi)\sin^{2}2\varphi}\text{.}\label{TWO}%
\end{equation}%
\begin{figure}[t]
\begin{center}
\includegraphics{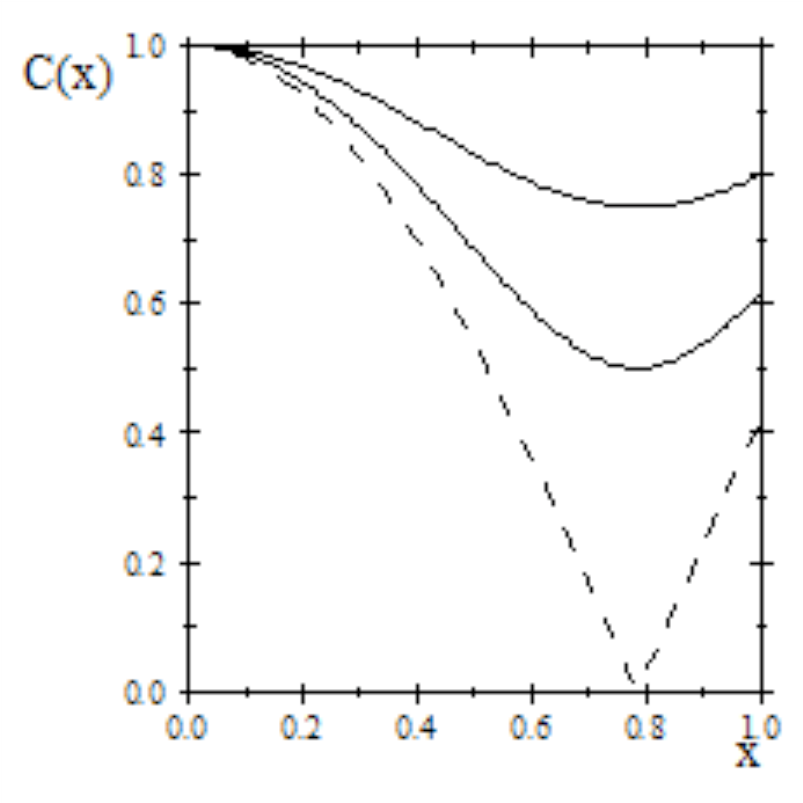}
\caption{$\emph{C}_{\text{momentum}}^{\left(  S^{\prime}\right)  }\left(
\varphi\right)  $ vs. $\varphi$ for $x\equiv\varphi\in\left[  0\text{,
}1\right]  :$ $\xi=\frac{1}{2}$ (dash), $\xi=\frac{1}{4}$ (thin solid),
$\xi=\frac{1}{8}$ (thick solid).}%
\end{center}
\end{figure}
Let us discuss the main consequences that originate from Eqs. (\ref{ONE}) and
(\ref{TWO}). Observe that both the $ss$ and $mm$-entanglements in the moving
inertial reference frame $S^{\prime}$ are functions of either the entangling
parameter $\xi$ or the angle of rotation $\varphi$. This implies that the
analysis of the monotonic behavior of the entanglements-changes in different
reference frames is tricky since it involves changes to both $\xi$ and
$\varphi$. From (\ref{ONE}) it follows that for a given value of the rotation
angle $\varphi$, the $ss$-entanglement is an \emph{increasing} function of
$\xi$ for $0<\xi<\frac{1}{2}$ while it is a \emph{decreasing} function of
$\xi$ for $\frac{1}{2}<\xi<1$ (see Figure $3$). Furthermore, for fixed values
of $\xi$, Lorentz boosts with greater angles of rotation exhibit an
attenuating power on the $ss$-entanglement that is stronger than that related
to boosts with smaller angles (see Figure $3$). From (\ref{TWO}) it follows
that, unlike the $ss$-entanglement, for a given value of the rotation angle
$\varphi$, the $mm$-entanglement is a \emph{decreasing} function of $\xi$ for
$0<\xi<\frac{1}{2}$ while it is an \emph{increasing} function of $\xi$ for
$\frac{1}{2}<\xi<1$ (see Figure $4$). However, like the $ss$-entanglement, for
fixed values of $\xi$, Lorentz boosts with greater angles of rotation exhibit
an attenuating power on the $mm$-entanglement that is stronger than that
related to boosts with smaller angles (see Figure $4$). For the sake of
clarity and without loss of relevant information, we focus our attention in
the rest of the manuscript on values of the rotation angle $\varphi$ in the
interval $\left[  0\text{, }1\right]  $. From (\ref{ONE}) it also turns out
that for a given value of the entangling parameter $\xi$, the $ss$%
-entanglement is a \emph{decreasing }function of $\varphi$ for $0<\varphi
<\frac{\pi}{4}$ while it is an \emph{increasing }function of $\varphi$ for
$\frac{\pi}{4}<\varphi<1$ (see Figure $5$). In particular, unlike the analysis
presented in \cite{suda}, we are able to show from (\ref{ONE}) that the
absolute value of the rate of change in $\varphi$ of $\emph{C}_{\text{spin}%
}^{\left(  S^{\prime}\right)  }\left(  \varphi\right)  $ (that is, its
"speed") depends on the degree of entanglement of the initial state in the
rest frame: the spin-entanglement degradation occurs at a faster rate for
states with a higher degree of entanglement,%
\begin{equation}
\left\vert \left(  \frac{d\emph{C}_{\text{spin}}^{\left(  S^{\prime}\right)
}\left(  \varphi\right)  }{d\varphi}\right)  _{\xi=\xi_{M}}\right\vert
\geq\left\vert \left(  \frac{d\emph{C}_{\text{spin}}^{\left(  S^{\prime
}\right)  }\left(  \varphi\right)  }{d\varphi}\right)  _{\xi=\xi_{m}%
}\right\vert \text{,}%
\end{equation}
with $\xi_{M}\geq\xi_{m}$. Furthermore, for fixed values of $\varphi$, to
\emph{lower} values of $\xi$ correspond \emph{lower} values of the
$ss$-entanglement (see Figure $5$). From (\ref{TWO}) it follows that, like for
the $ss$-entanglement, for a given value of the entangling parameter $\xi$,
the $mm$-entanglement is a \emph{decreasing }function of $\varphi$ for
$0<\varphi<\frac{\pi}{4}$ while it is an \emph{increasing }function of
$\varphi$ for $\frac{\pi}{4}<\varphi<1$ (see Figure $6$). However, unlike the
$ss$-entanglement, for fixed values of $\varphi$, to \emph{lower} values of
$\xi$ correspond \emph{higher} values of the $mm$-entanglement (see Figure
$6$). In particular, unlike the study appeared in \cite{suda}, we can state
from (\ref{TWO}) that the absolute value of the rate of change in $\varphi$ of
$\emph{C}_{\text{momentum}}^{\left(  S^{\prime}\right)  }\left(
\varphi\right)  $ depends on the degree of entanglement of the initial state
in the rest frame: the momentum-entanglement degradation occurs at a slower
rate for states with higher degree of entanglement,%
\begin{equation}
\left\vert \left(  \frac{d\emph{C}_{\text{momentum}}^{\left(  S^{\prime
}\right)  }\left(  \varphi\right)  }{d\varphi}\right)  _{\xi=\xi_{M}%
}\right\vert \leq\left\vert \left(  \frac{d\emph{C}_{\text{momentum}}^{\left(
S^{\prime}\right)  }\left(  \varphi\right)  }{d\varphi}\right)  _{\xi=\xi_{m}%
}\right\vert \text{,}%
\end{equation}
with $\xi_{M}\geq\xi_{m}$. We also stress that our analysis shows that no sum
of $ss$ and $mm$-entanglements is conserved in any pair of inertial reference
frames in relative motion for any value of the entangling parameter,%
\begin{equation}
\emph{C}_{\text{spin}}^{\left(  S\right)  }\left(  \xi\right)  +\emph{C}%
_{\text{momentum}}^{\left(  S\right)  }\neq\emph{C}_{\text{spin}}^{\left(
S^{\prime}\right)  }\left(  \xi\text{, }\varphi\right)  +\emph{C}%
_{\text{momentum}}^{\left(  S^{\prime}\right)  }\left(  \xi\text{, }%
\varphi\right)  \text{.}%
\end{equation}
Finally, it turns out that both $\xi$-changes and $\varphi$-changes of
variations of either $ss$ or $mm$-entanglements, $\Delta\emph{C}_{\text{spin}%
}$ and $\Delta\emph{C}_{\text{momentum}}$, respectively, exhibit the same
monotonic behavior on the permitted range of values for each parameter. More
explicitly, it follows that%
\begin{equation}
\emph{sign}\left[  \left(  \frac{\partial\Delta\emph{C}_{\text{spin}}%
}{\partial\xi}\right)  _{\varphi}\right]  =\emph{sign}\left[  \left(
\frac{\partial\Delta\emph{C}_{\text{momentum}}}{\partial\xi}\right)
_{\varphi}\right]  \text{,}\label{O1}%
\end{equation}
and,%
\begin{equation}
\emph{sign}\left[  \left(  \frac{\partial\Delta\emph{C}_{\text{spin}}%
}{\partial\varphi}\right)  _{\xi}\right]  =\emph{sign}\left[  \left(
\frac{\partial\Delta\emph{C}_{\text{momentum}}}{\partial\varphi}\right)
_{\xi}\right]  \text{,}\label{O2}%
\end{equation}
where the variations $\Delta\emph{C}_{\text{spin}}$ and $\Delta\emph{C}%
_{\text{momentum}}$ are defined as,%
\begin{equation}
\Delta\emph{C}_{\text{spin}}\left(  \xi\text{, }\varphi\right)  \overset
{\text{def}}{=}\emph{C}_{\text{spin}}^{\left(  S\right)  }\left(  \xi\right)
-\emph{C}_{\text{spin}}^{\left(  S^{\prime}\right)  }\left(  \xi\text{,
}\varphi\right)  \text{,}%
\end{equation}
and,%
\begin{equation}
\Delta\emph{C}_{\text{momentum}}\left(  \xi\text{, }\varphi\right)
\overset{\text{def}}{=}\emph{C}_{\text{momentum}}^{\left(  S\right)
}-\emph{C}_{\text{momentum}}^{\left(  S^{\prime}\right)  }\left(  \xi\text{,
}\varphi\right)  \text{,}%
\end{equation}
respectively. For an illustrative justification of Eqs. (\ref{O1}) and
(\ref{O2}), see Figures $7$ and $8$, respectively. This last finding implies
that no entanglement compensation between spins and momenta occurs and neither
it is required by the Lorentz invariance of the joint entanglement of the
entire wave-function since no analytical constraint exists between this
quantity and the $ss$ and $mm$-entanglements.
\begin{figure}[t]
\begin{center}
\includegraphics{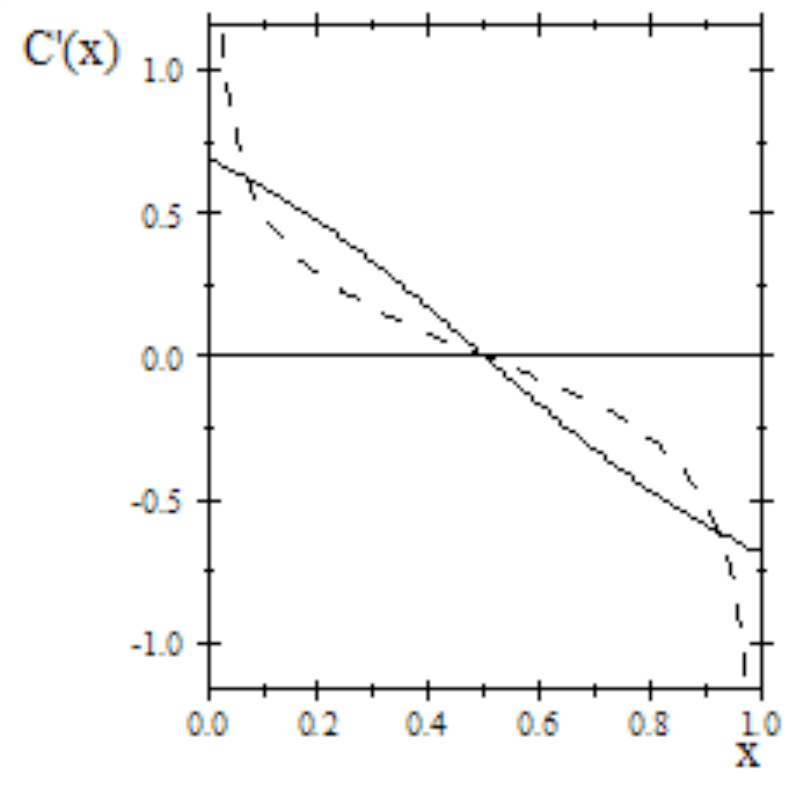}
\caption{$\frac{\partial\Delta C_{\text{spin}}\left(  \xi\text{, }\bar
{\varphi}\right)  }{\partial\xi}$ (dash) and $\frac{\partial\Delta
C_{\text{momentum}}\left(  \xi\text{, }\bar{\varphi}\right)  }{\partial\xi}$
(solid) with $x\equiv\xi$ vs. $\xi$ for $\bar{\varphi}=\frac{\pi}{10}$.}%
\end{center}
\end{figure}

\section{Final Remarks}

In this article, we have analyzed special\textbf{ }relativistic effects on the
entanglement between either spins or momenta of composite quantum systems of
two spin-$\frac{1}{2}$ massive particles, either distinguishable or
indistinguishable, in inertial reference frames in relative motion.

\emph{Indistinguishable particles with balanced (but not maximal) entanglement
configuration}. For the case of indistinguishable particles, we consider a
single balanced scenario where the momenta of the pair are well-defined but
not maximally entangled in the rest frame while the spins of the pair are
described by a one-parameter ($\eta$) family of entangled bipartite states
(see Eq. (\ref{as})). We find out that in any scenario neither the $ss$ nor
the $mm$ entanglements are Lorentz invariant quantities (see Eqs.
(\ref{nota3}) and (\ref{nota4})). In particular, for any value of the
entangling parameters, both $ss$ and $mm$-entanglements are attenuated by
Lorentz transformations. Moreover, the change in entanglement for the momenta
is the same as the change in entanglement for spins (see Eq. (\ref{same1}) and
Figures $1$ and $2$). In particular (see Eq. (\ref{sopra})), when $\eta=0$ or
$\eta=1$, we recover the main result appeared in {\cite{suda}.}

\emph{Distinguishable particles with unbalanced entanglement configuration}.
For the case of distinguishable particles, we consider an unbalanced scenario
where the momenta of the pair are well-defined and maximally entangled in the
rest frame while the spins of the pair are described by a one-parameter ($\xi
$) family of non-maximally entangled bipartite states (see (\ref{as0})). We
present an extensive investigation of the behavior of two-parameters (the
angle of rotation $\varphi$ and the entangling parameter $\xi$) $ss$ and
$mm$-entanglements for a two-particle quantum composite system of massive
fermions in inertial reference frames in relative motion. We show that in any
case neither the $ss$ nor the $mm$ entanglements quantified by means of
Wootters' concurrence were Lorentz invariant quantities (see Eqs. (\ref{ONE}),
(\ref{TWO})): the total amount of entanglement regarded as the sum of these
entanglements is not the same in all inertial frames. Furthermore, for any
value of the entangling parameter, both $ss$ and $mm$-entanglements are
attenuated by Lorentz transformations (see Figures $3$, $4$, $5$, $6$) and
their parametric rates of change with respect to the entanglements observed in
a rest frame have the same monotonic behavior (see Figures $7$ and $8$). In
particular, in Appendix D, we also consider an additional scenario of
distinguishable particles (see Eq. (\ref{qqq})) where in the limiting case of
$\xi=\frac{1}{2}$ and $m_{e}=m_{\mu}\equiv m$ (indistinguishable particles,
two identical electrons), the main result of \cite{suda} is reproduced.
However, unlike the main finding appeared in \cite{suda} and our generalized
result in (\ref{same1}) which holds for indistinguishable particles, we show
that in general the change in entanglement for the momenta is \emph{not} the
same as the change in entanglement for the spins when considering
distinguishable particles. Furthermore, unlike the main result presented in
\cite{adami}, we provide clear evidences that the rate of entanglement changes
in both parameters $\xi$ and $\varphi$ are in the same direction (not in the
opposite direction). Surprisingly, even allowing for non-maximal entanglement
in the spin degrees in the rest frame $S$ (while keeping the momentum
entanglement maximal in $S$), no entanglement compensation from the momentum
to spin entanglement occurs for any possible pair of inertial reference frames
one of which at rest.

In conclusion, although a thorough characterization of relativistic properties
of quantum entanglement is far from being achieved, our study provides general
enough results towards such direction.
\begin{figure}[t]
\begin{center}
\includegraphics{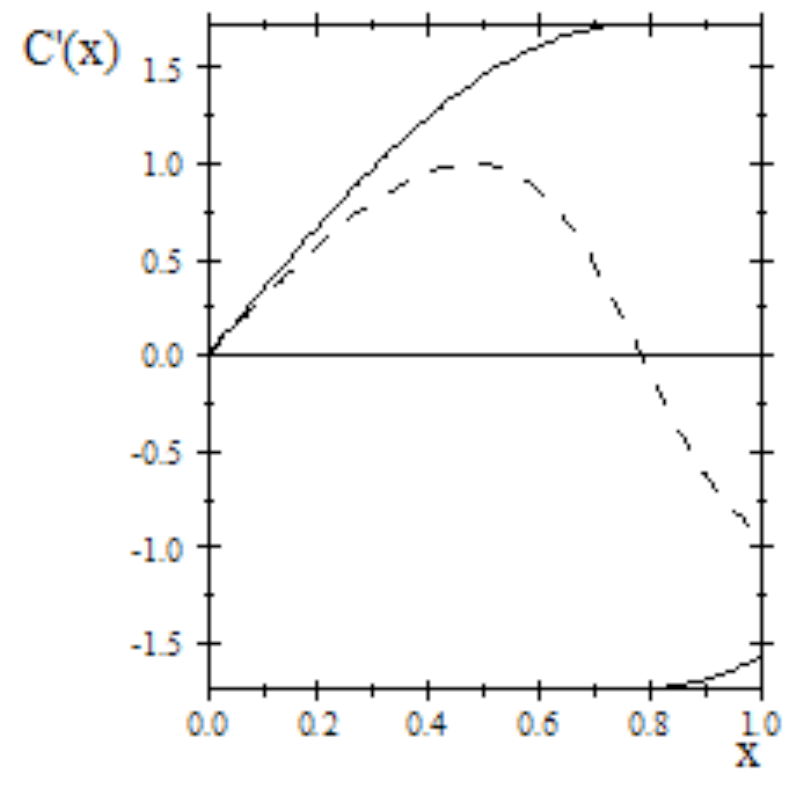}
\caption{$\frac{\partial\Delta C_{\text{spin}}\left(  \bar{\xi}\text{,
}\varphi\right)  }{\partial\varphi}$ (solid) and $\frac{\partial\Delta
C_{\text{momentum}}\left(  \bar{\xi}\text{, }\varphi\right)  }{\partial
\varphi}$ (dash) with $x\equiv\varphi$ vs. $\varphi$ for $\bar{\xi}=\frac
{1}{4}$.}%
\end{center}
\end{figure}

\begin{acknowledgments}
The research of C. Cafaro and S. Mancini has received funding from the
European Commission's Seventh Framework Programme (FP7/2007--2013) under grant
agreements no. 213681.
\end{acknowledgments}

\appendix

\section{Decomposition of density operators in terms of Pauli matrices}

In this Appendix, we derive Eq. (\ref{ne}). Consider a two-qubit system with
Hilbert space $\mathcal{H}=
\mathbb{C}
^{2}\otimes
\mathbb{C}
^{2}$ and computational basis $\mathcal{B}_{\mathcal{H}}=\left\{  \left\vert
00\right\rangle \text{, }\left\vert 01\right\rangle \text{, }\left\vert
10\right\rangle \text{, }\left\vert 11\right\rangle \right\}  $. It can be
shown that a general two-qubit state $\hat{\rho}$ can always be written up to
local unitary equivalence to a state of the following form \cite{luo},%
\begin{equation}
\hat{\rho}=\frac{1}{4}\left[  I_{4\times4}+\vec{a}\cdot\vec{\sigma}\otimes
I_{2\times2}+I_{2\times2}\otimes\vec{b}\cdot\vec{\sigma}+\sum_{i=1}^{3}%
c_{i}\sigma_{i}\otimes\sigma_{i}\right]  \text{,} \label{parametrization1}%
\end{equation}
that is, after some algebra,%
\begin{equation}
\hat{\rho}=\frac{1}{4}\left(
\begin{array}
[c]{cccc}%
a_{3}+b_{3}+c_{3}+1 & b_{1}-ib_{2} & a_{1}-ia_{2} & c_{1}-c_{2}\\
b_{1}+ib_{2} & a_{3}-b_{3}-c_{3}+1 & c_{1}+c_{2} & a_{1}-ia_{2}\\
a_{1}+ia_{2} & c_{1}+c_{2} & b_{3}-a_{3}-c_{3}+1 & b_{1}-ib_{2}\\
c_{1}-c_{2} & a_{1}+ia_{2} & b_{1}+ib_{2} & c_{3}-b_{3}-a_{3}+1
\end{array}
\right)  \text{,} \label{rogeneral}%
\end{equation}
where $\vec{a}$ and $\vec{b}\in
\mathbb{R}
^{3}$ are given by $\vec{a}=\left(  a_{1}\text{, }a_{2}\text{, }a_{3}\right)
$, $\vec{b}=\left(  b_{1}\text{, }b_{2}\text{, }b_{3}\right)  $ and
$\vec{\sigma}=\left(  \sigma_{x}\text{, }\sigma_{y}\text{, }\sigma_{z}\right)
$ is the operator vector of Pauli matrices. Equating $\hat{\rho}_{\psi_{+}%
}\left(  \xi\right)  $ in (\ref{neo}) to (\ref{rogeneral}), we obtain%
\begin{equation}
a_{1}=a_{2}=b_{1}=b_{2}=0\text{, }c_{1}=c_{2}=2\sqrt{\xi\left(  1-\xi\right)
}\text{, }a_{3}=1-2\xi\text{, }b_{3}=-\left(  1-2\xi\right)  \text{, }%
c_{3}=-1\text{.}%
\end{equation}
Therefore, in terms of Pauli matrices, $\hat{\rho}_{\psi_{+}}\left(
\xi\right)  $ becomes%
\begin{equation}
\hat{\rho}_{\psi_{+}}\left(  \xi\right)  =\frac{1}{4}\left[
\begin{array}
[c]{c}%
I_{2\times2}^{\left(  1\right)  }\otimes I_{2\times2}^{\left(  2\right)
}+\left(  1-2\xi\right)  \sigma_{z}^{\left(  1\right)  }\otimes I_{2\times
2}^{\left(  2\right)  }-\left(  1-2\xi\right)  I_{2\times2}^{\left(  1\right)
}\otimes\sigma_{z}^{\left(  2\right)  }+2\sqrt{\xi\left(  1-\xi\right)
}\sigma_{x}^{\left(  1\right)  }\otimes\sigma_{x}^{\left(  2\right)  }+\\
\\
+2\sqrt{\xi\left(  1-\xi\right)  }\sigma_{y}^{\left(  1\right)  }\otimes
\sigma_{y}^{\left(  2\right)  }-\sigma_{z}^{\left(  1\right)  }\otimes
\sigma_{z}^{\left(  2\right)  }%
\end{array}
\right]  \text{.}%
\end{equation}
Observe that for $\xi=\frac{1}{2}$ we obtain,%
\begin{equation}
\hat{\rho}_{\psi_{+}}\left(  \frac{1}{2}\right)  =\frac{1}{4}\left[
I_{2\times2}^{\left(  1\right)  }\otimes I_{2\times2}^{\left(  2\right)
}+\sigma_{x}^{\left(  1\right)  }\otimes\sigma_{x}^{\left(  2\right)  }%
+\sigma_{y}^{\left(  1\right)  }\otimes\sigma_{y}^{\left(  2\right)  }%
-\sigma_{z}^{\left(  1\right)  }\otimes\sigma_{z}^{\left(  2\right)  }\right]
\text{.}%
\end{equation}
Similarly, equating $\hat{\rho}_{\psi_{-}}\left(  \xi\right)  $ in (\ref{neo})
to (\ref{rogeneral}), we get%
\begin{equation}
a_{1}=a_{2}=b_{1}=b_{2}=0\text{, }c_{1}=c_{2}=-2\sqrt{\xi\left(  1-\xi\right)
}\text{, }a_{3}=1-2\xi\text{, }b_{3}=-\left(  1-2\xi\right)  \text{, }%
c_{3}=-1\text{.}%
\end{equation}
In terms of Pauli matrices, $\hat{\rho}_{\psi_{-}}\left(  \xi\right)  $ reads,%
\begin{equation}
\hat{\rho}_{\psi_{-}}\left(  \xi\right)  =\frac{1}{4}\left[
\begin{array}
[c]{c}%
I_{2\times2}^{\left(  1\right)  }\otimes I_{2\times2}^{\left(  2\right)
}+\left(  1-2\xi\right)  \sigma_{z}^{\left(  1\right)  }\otimes I_{2\times
2}^{\left(  2\right)  }-\left(  1-2\xi\right)  I_{2\times2}^{\left(  1\right)
}\otimes\sigma_{z}^{\left(  2\right)  }-2\sqrt{\xi\left(  1-\xi\right)
}\sigma_{x}^{\left(  1\right)  }\otimes\sigma_{x}^{\left(  2\right)  }+\\
\\
-2\sqrt{\xi\left(  1-\xi\right)  }\sigma_{y}^{\left(  1\right)  }\otimes
\sigma_{y}^{\left(  2\right)  }-\sigma_{z}^{\left(  1\right)  }\otimes
\sigma_{z}^{\left(  2\right)  }%
\end{array}
\right]  \text{,}%
\end{equation}
and for $\xi=\frac{1}{2}$ it becomes,%
\begin{equation}
\hat{\rho}_{\psi_{-}}\left(  \frac{1}{2}\right)  =\frac{1}{4}\left[
I_{2\times2}^{\left(  1\right)  }\otimes I_{2\times2}^{\left(  2\right)
}-\sigma_{x}^{\left(  1\right)  }\otimes\sigma_{x}^{\left(  2\right)  }%
-\sigma_{y}^{\left(  1\right)  }\otimes\sigma_{y}^{\left(  2\right)  }%
-\sigma_{z}^{\left(  1\right)  }\otimes\sigma_{z}^{\left(  2\right)  }\right]
\text{.}%
\end{equation}
In summary, we have shown that%
\begin{equation}
\hat{\rho}_{\psi\pm}\left(  \xi\right)  =\frac{1}{4}\left[
\begin{array}
[c]{c}%
I_{2\times2}^{\left(  1\right)  }\otimes I_{2\times2}^{\left(  2\right)
}+\left(  1-2\xi\right)  \sigma_{z}^{\left(  1\right)  }\otimes I_{2\times
2}^{\left(  2\right)  }-\left(  1-2\xi\right)  I_{2\times2}^{\left(  1\right)
}\otimes\sigma_{z}^{\left(  2\right)  }\pm2\sqrt{\xi\left(  1-\xi\right)
}\sigma_{x}^{\left(  1\right)  }\otimes\sigma_{x}^{\left(  2\right)  }+\\
\\
\pm2\sqrt{\xi\left(  1-\xi\right)  }\sigma_{y}^{\left(  1\right)  }%
\otimes\sigma_{y}^{\left(  2\right)  }-\sigma_{z}^{\left(  1\right)  }%
\otimes\sigma_{z}^{\left(  2\right)  }%
\end{array}
\right]  \text{,}%
\end{equation}
and in the limiting case of $\xi=\frac{1}{2}$ we recover the case of maximally
entangled pure Bell states.

\section{Lorentz transformation of the spin density operator}

In this Appendix, we derive Eq. (\ref{nue}). Substituting (\ref{b}) and
(\ref{c}) into (\ref{t-spin}), we get%
\begin{equation}
\hat{\rho}_{\text{spin}}^{\prime}=\frac{1}{2}\mathcal{D}_{A}\left(  p_{A_{1}%
}\right)  \mathcal{D}_{B}\left(  p_{B_{1}}\right)  \hat{\rho}_{\text{spin}%
}\mathcal{D}_{A}^{\dagger}\left(  p_{A_{1}}\right)  \mathcal{D}_{B}^{\dagger
}\left(  p_{B_{1}}\right)  +\frac{1}{2}\mathcal{D}_{A}\left(  p_{A_{2}%
}\right)  \mathcal{D}_{B}\left(  p_{B_{2}}\right)  \hat{\rho}_{\text{spin}%
}\mathcal{D}_{A}^{\dagger}\left(  p_{A_{2}}\right)  \mathcal{D}_{B}^{\dagger
}\left(  p_{B_{2}}\right)  \text{,} \label{53}%
\end{equation}
where $\hat{\rho}_{\text{spin}}$ equals $\hat{\rho}_{\psi\pm}\left(
\xi\right)  $ in (\ref{ne}). Using (\ref{D}), we observe that the following
transformation rules hold%
\begin{align}
\mathcal{D}_{A}\left(  p_{A_{1}}\right)  \sigma_{x}^{\left(  A\right)
}\mathcal{D}_{A}^{\dagger}\left(  p_{A_{1}}\right)   &  =\left(  \cos
\frac{\varphi}{2}-i\sin\frac{\varphi}{2}\sigma_{z}^{\left(  A\right)
}\right)  \sigma_{x}^{\left(  A\right)  }\left(  \cos\frac{\varphi}{2}%
+i\sin\frac{\varphi}{2}\sigma_{z}^{\left(  A\right)  }\right)  =\cos
\varphi\sigma_{x}^{\left(  A\right)  }+\sin\varphi\sigma_{y}^{\left(
A\right)  }\text{,}\nonumber\\
& \nonumber\\
\mathcal{D}_{B}\left(  p_{B_{1}}\right)  \sigma_{x}^{\left(  B\right)
}\mathcal{D}_{B}^{\dagger}\left(  p_{B_{1}}\right)   &  =\left(  \cos
\frac{\varphi}{2}+i\sin\frac{\varphi}{2}\sigma_{z}^{\left(  B\right)
}\right)  \sigma_{x}^{\left(  B\right)  }\left(  \cos\frac{\varphi}{2}%
-i\sin\frac{\varphi}{2}\sigma_{z}^{\left(  B\right)  }\right)  =\cos
\varphi\sigma_{x}^{\left(  B\right)  }-\sin\varphi\sigma_{y}^{\left(
B\right)  }\text{,}\nonumber\\
& \nonumber\\
\mathcal{D}_{A}\left(  p_{A_{1}}\right)  \sigma_{y}^{\left(  A\right)
}\mathcal{D}_{A}^{\dagger}\left(  p_{A_{1}}\right)   &  =\left(  \cos
\frac{\varphi}{2}-i\sin\frac{\varphi}{2}\sigma_{z}^{\left(  A\right)
}\right)  \sigma_{y}^{\left(  A\right)  }\left(  \cos\frac{\varphi}{2}%
+i\sin\frac{\varphi}{2}\sigma_{z}^{\left(  A\right)  }\right)  =-\sin
\varphi\sigma_{x}^{\left(  A\right)  }+\cos\varphi\sigma_{y}^{\left(
A\right)  }\text{,}\nonumber\\
& \nonumber\\
\mathcal{D}_{B}\left(  p_{B_{1}}\right)  \sigma_{y}^{\left(  B\right)
}\mathcal{D}_{B}^{\dagger}\left(  p_{B_{1}}\right)   &  =\left(  \cos
\frac{\varphi}{2}+i\sin\frac{\varphi}{2}\sigma_{z}^{\left(  B\right)
}\right)  \sigma_{y}^{\left(  B\right)  }\left(  \cos\frac{\varphi}{2}%
-i\sin\frac{\varphi}{2}\sigma_{z}^{\left(  B\right)  }\right)  =\sin
\varphi\sigma_{x}^{\left(  B\right)  }+\cos\varphi\sigma_{y}^{\left(
B\right)  }\text{,} \label{51}%
\end{align}
and,%
\begin{align}
\mathcal{D}_{A}\left(  p_{A_{2}}\right)  \sigma_{x}^{\left(  A\right)
}\mathcal{D}_{A}^{\dagger}\left(  p_{A_{2}}\right)   &  =\left(  \cos
\frac{\varphi}{2}+i\sin\frac{\varphi}{2}\sigma_{z}^{\left(  A\right)
}\right)  \sigma_{x}^{\left(  A\right)  }\left(  \cos\frac{\varphi}{2}%
-i\sin\frac{\varphi}{2}\sigma_{z}^{\left(  A\right)  }\right)  =\cos
\varphi\sigma_{x}^{\left(  A\right)  }-\sin\varphi\sigma_{y}^{\left(
A\right)  }\text{,}\nonumber\\
& \nonumber\\
\mathcal{D}_{B}\left(  p_{B_{2}}\right)  \sigma_{x}^{\left(  B\right)
}\mathcal{D}_{B}^{\dagger}\left(  p_{B_{2}}\right)   &  =\left(  \cos
\frac{\varphi}{2}-i\sin\frac{\varphi}{2}\sigma_{z}^{\left(  B\right)
}\right)  \sigma_{x}^{\left(  B\right)  }\left(  \cos\frac{\varphi}{2}%
+i\sin\frac{\varphi}{2}\sigma_{z}^{\left(  B\right)  }\right)  =\cos
\varphi\sigma_{x}^{\left(  B\right)  }+\sin\varphi\sigma_{y}^{\left(
B\right)  }\text{,}\nonumber\\
& \nonumber\\
\mathcal{D}_{A}\left(  p_{A_{2}}\right)  \sigma_{x}^{\left(  A\right)
}\mathcal{D}_{A}^{\dagger}\left(  p_{A_{2}}\right)   &  =\left(  \cos
\frac{\varphi}{2}+i\sin\frac{\varphi}{2}\sigma_{z}^{\left(  A\right)
}\right)  \sigma_{y}^{\left(  A\right)  }\left(  \cos\frac{\varphi}{2}%
-i\sin\frac{\varphi}{2}\sigma_{z}^{\left(  A\right)  }\right)  =\sin
\varphi\sigma_{x}^{\left(  A\right)  }+\cos\varphi\sigma_{y}^{\left(
A\right)  }\text{,}\nonumber\\
& \nonumber\\
\mathcal{D}_{B}\left(  p_{B_{2}}\right)  \sigma_{x}^{\left(  B\right)
}\mathcal{D}_{B}^{\dagger}\left(  p_{B_{2}}\right)   &  =\left(  \cos
\frac{\varphi}{2}-i\sin\frac{\varphi}{2}\sigma_{z}^{\left(  B\right)
}\right)  \sigma_{y}^{\left(  B\right)  }\left(  \cos\frac{\varphi}{2}%
+i\sin\frac{\varphi}{2}\sigma_{z}^{\left(  B\right)  }\right)  =-\sin
\varphi\sigma_{x}^{\left(  B\right)  }+\cos\varphi\sigma_{y}^{\left(
B\right)  }\text{.} \label{52}%
\end{align}
Therefore, using (\ref{51}) and (\ref{52}), after some tedious algebra it
follows that $\hat{\rho}_{\text{spin}}^{\prime}$ in (\ref{53}) becomes%
\begin{equation}
\hat{\rho}_{\text{spin}}^{\prime}=\frac{1}{4}\left[
\begin{array}
[c]{c}%
I_{2\times2}^{\left(  A\right)  }\otimes I_{2\times2}^{\left(  B\right)
}+\left(  1-2\xi\right)  \sigma_{z}^{\left(  A\right)  }\otimes I_{2\times
2}^{\left(  B\right)  }-\left(  1-2\xi\right)  I_{2\times2}^{\left(  A\right)
}\otimes\sigma_{z}^{\left(  B\right)  }\pm2\sqrt{\xi\left(  1-\xi\right)
}\cos2\varphi\sigma_{x}^{\left(  A\right)  }\otimes\sigma_{x}^{\left(
B\right)  }+\\
\\
\pm2\sqrt{\xi\left(  1-\xi\right)  }\cos2\varphi\sigma_{y}^{\left(  A\right)
}\otimes\sigma_{y}^{\left(  B\right)  }-\sigma_{z}^{\left(  A\right)  }%
\otimes\sigma_{z}^{\left(  B\right)  }\text{.}%
\end{array}
\right]  \text{,}%
\end{equation}
that is, $\hat{\rho}_{\text{spin}}^{\prime}$ equals $\hat{\rho}_{\psi\pm
}^{\left(  \Lambda\right)  }\left(  \xi\text{, }\varphi\right)  $ in
(\ref{nue}).

\section{Lorentz transformation of the momentum density operator}

In this Appendix, we derive Eq. (\ref{derivation}). Using Eqs. (\ref{pp1}),
(\ref{pp22})\ and (\ref{pp3}), it follows that%
\begin{align}
\left\vert \Lambda\vec{p}_{A_{1}}\right\rangle \left\langle \Lambda\vec
{p}_{A_{1}}\right\vert  &  =\frac{I_{2\times2}^{\left(  A\right)  }%
+\tilde{\sigma}_{z}^{\left(  A\right)  }}{2}\text{, }\left\vert \Lambda\vec
{p}_{B_{1}}\right\rangle \left\langle \Lambda\vec{p}_{B_{1}}\right\vert
=\frac{I_{2\times2}^{\left(  B\right)  }+\tilde{\sigma}_{z}^{\left(  B\right)
}}{2}\text{, }\nonumber\\
& \nonumber\\
\left\vert \Lambda\vec{p}_{A_{2}}\right\rangle \left\langle \Lambda\vec
{p}_{A_{2}}\right\vert  &  =\frac{I_{2\times2}^{\left(  A\right)  }%
-\tilde{\sigma}_{z}^{\left(  A\right)  }}{2}\text{, }\left\vert \Lambda\vec
{p}_{B_{2}}\right\rangle \left\langle \Lambda\vec{p}_{B_{2}}\right\vert
=\frac{I_{2\times2}^{\left(  B\right)  }-\tilde{\sigma}_{z}^{\left(  B\right)
}}{2}\text{,} \label{z1}%
\end{align}
and,%
\begin{align}
\left\vert \Lambda\vec{p}_{A_{1}}\right\rangle \left\langle \Lambda\vec
{p}_{A_{2}}\right\vert  &  =\frac{\tilde{\sigma}_{x}^{\left(  A\right)
}+i\tilde{\sigma}_{y}^{\left(  A\right)  }}{2}\text{, }\left\vert \Lambda
\vec{p}_{A_{2}}\right\rangle \left\langle \Lambda\vec{p}_{A_{1}}\right\vert
=\frac{\tilde{\sigma}_{x}^{\left(  A\right)  }-i\tilde{\sigma}_{y}^{\left(
A\right)  }}{2}\text{,}\nonumber\\
& \nonumber\\
\left\vert \Lambda\vec{p}_{B_{1}}\right\rangle \left\langle \Lambda\vec
{p}_{B_{2}}\right\vert  &  =\frac{\tilde{\sigma}_{x}^{\left(  B\right)
}+i\tilde{\sigma}_{y}^{\left(  B\right)  }}{2}\text{, }\left\vert \Lambda
\vec{p}_{B_{2}}\right\rangle \left\langle \Lambda\vec{p}_{B_{1}}\right\vert
=\frac{\tilde{\sigma}_{x}^{\left(  B\right)  }-i\tilde{\sigma}_{y}^{\left(
B\right)  }}{2}\text{.} \label{z2}%
\end{align}
Using (\ref{z1})\ and (\ref{z2}), $\hat{\rho}_{\text{momentum}}^{\prime}$ in
(\ref{derivation}) becomes%
\begin{equation}
\hat{\rho}_{\text{momentum}}^{\prime}=\frac{1}{2}\left[
\begin{array}
[c]{c}%
\left\langle 1|1\right\rangle \left(  \frac{I_{2\times2}^{\left(  A\right)
}+\tilde{\sigma}_{z}^{\left(  A\right)  }}{2}\right)  \otimes\left(
\frac{I_{2\times2}^{\left(  B\right)  }+\tilde{\sigma}_{z}^{\left(  B\right)
}}{2}\right)  +\left\langle 2|1\right\rangle \left(  \frac{\tilde{\sigma}%
_{x}^{\left(  A\right)  }+i\tilde{\sigma}_{y}^{\left(  A\right)  }}{2}\right)
\otimes\left(  \frac{\tilde{\sigma}_{x}^{\left(  B\right)  }+i\tilde{\sigma
}_{y}^{\left(  B\right)  }}{2}\right)  +\\
\\
+\left\langle 1|2\right\rangle \left(  \frac{\tilde{\sigma}_{x}^{\left(
A\right)  }-i\tilde{\sigma}_{y}^{\left(  A\right)  }}{2}\right)
\otimes\left(  \frac{\tilde{\sigma}_{x}^{\left(  B\right)  }-i\tilde{\sigma
}_{y}^{\left(  B\right)  }}{2}\right)  +\left\langle 2|2\right\rangle \left(
\frac{I_{2\times2}^{\left(  A\right)  }-\tilde{\sigma}_{z}^{\left(  A\right)
}}{2}\right)  \otimes\left(  \frac{I_{2\times2}^{\left(  B\right)  }%
-\tilde{\sigma}_{z}^{\left(  B\right)  }}{2}\right)
\end{array}
\right]  \text{.} \label{23}%
\end{equation}
Furthermore, substituting (\ref{47}) into (\ref{b})\ and (\ref{c}), it follows
that $\left\langle 1|1\right\rangle =\left\langle 2|2\right\rangle =1$ while
$\left\langle 1|2\right\rangle $ and $\left\langle 2|1\right\rangle $ are
given by,%
\begin{equation}
\left\langle 1|2\right\rangle =\text{Tr}\left(
\begin{array}
[c]{cccc}%
0 & 0 & 0 & 0\\
0 & \left(  1-\xi\right)  \left(  \cos2\varphi+i\sin2\varphi\right)  &
\pm\sqrt{\xi\left(  1-\xi\right)  }\left(  \cos2\varphi+i\sin2\varphi\right)
& 0\\
0 & \pm\sqrt{\xi\left(  1-\xi\right)  }\left(  \cos2\varphi-i\sin
2\varphi\right)  & \xi\left(  \cos2\varphi-i\sin2\varphi\right)  & 0\\
0 & 0 & 0 & 0
\end{array}
\right)  =\cos2\varphi-i\left(  2\xi-1\right)  \sin2\varphi\text{,} \label{12}%
\end{equation}
and,%
\begin{equation}
\left\langle 2|1\right\rangle =\text{Tr}\left(
\begin{array}
[c]{cccc}%
0 & 0 & 0 & 0\\
0 & \left(  1-\xi\right)  \left(  \cos2\varphi-i\sin2\varphi\right)  &
\pm\sqrt{\xi\left(  1-\xi\right)  }\left(  \cos2\varphi-i\sin2\varphi\right)
& 0\\
0 & \pm\sqrt{\xi\left(  1-\xi\right)  }\left(  \cos2\varphi+i\sin
2\varphi\right)  & \xi\left(  \cos2\varphi+i\sin2\varphi\right)  & 0\\
0 & 0 & 0 & 0
\end{array}
\right)  =\cos2\varphi+i\left(  2\xi-1\right)  \sin2\varphi\text{,} \label{21}%
\end{equation}
respectively. Thus, substituting (\ref{12}) and (\ref{21}) into (\ref{23}), we
obtain%
\begin{equation}
\hat{\rho}_{\text{momentum}}^{\prime}=\frac{1}{2}\left[
\begin{array}
[c]{c}%
\left(  \frac{I_{2\times2}^{\left(  A\right)  }+\tilde{\sigma}_{z}^{\left(
A\right)  }}{2}\right)  \otimes\left(  \frac{I_{2\times2}^{\left(  B\right)
}+\tilde{\sigma}_{z}^{\left(  B\right)  }}{2}\right)  +\left[  \cos
2\varphi+i\left(  2\xi-1\right)  \sin2\varphi\right]  \left(  \frac
{\tilde{\sigma}_{x}^{\left(  A\right)  }+i\tilde{\sigma}_{y}^{\left(
A\right)  }}{2}\right)  \otimes\left(  \frac{\tilde{\sigma}_{x}^{\left(
B\right)  }+i\tilde{\sigma}_{y}^{\left(  B\right)  }}{2}\right)  +\\
\\
+\left[  \cos2\varphi-i\left(  2\xi-1\right)  \sin2\varphi\right]  \left(
\frac{\tilde{\sigma}_{x}^{\left(  A\right)  }-i\tilde{\sigma}_{y}^{\left(
A\right)  }}{2}\right)  \otimes\left(  \frac{\tilde{\sigma}_{x}^{\left(
B\right)  }-i\tilde{\sigma}_{y}^{\left(  B\right)  }}{2}\right)  +\left(
\frac{I_{2\times2}^{\left(  A\right)  }-\tilde{\sigma}_{z}^{\left(  A\right)
}}{2}\right)  \otimes\left(  \frac{I_{2\times2}^{\left(  B\right)  }%
-\tilde{\sigma}_{z}^{\left(  B\right)  }}{2}\right)
\end{array}
\right]  \text{.}%
\end{equation}
After some algebra, $\hat{\rho}_{\text{momentum}}^{\prime}$ may be finally
rewritten as $\hat{\rho}_{\text{momentum}}^{\prime}\left(  \xi\text{, }%
\varphi\right)  $ in (\ref{derivation}),%
\begin{equation}
\hat{\rho}_{\text{momentum}}^{\prime}\equiv\hat{\rho}_{\text{momentum}%
}^{\prime}\left(  \xi\text{, }\varphi\right)  =\frac{1}{4}\left[
\begin{array}
[c]{c}%
I_{2\times2}^{\left(  A\right)  }\otimes I_{2\times2}^{\left(  B\right)
}+\cos2\varphi\left(  \tilde{\sigma}_{x}^{\left(  A\right)  }\otimes
\tilde{\sigma}_{x}^{\left(  B\right)  }-\tilde{\sigma}_{y}^{\left(  A\right)
}\otimes\tilde{\sigma}_{y}^{\left(  B\right)  }\right)  +\tilde{\sigma}%
_{z}^{\left(  A\right)  }\otimes\tilde{\sigma}_{z}^{\left(  B\right)  }+\\
\\
+\left(  1-2\xi\right)  \sin2\varphi\left(  \tilde{\sigma}_{x}^{\left(
A\right)  }\otimes\tilde{\sigma}_{y}^{\left(  B\right)  }+\tilde{\sigma}%
_{y}^{\left(  A\right)  }\otimes\tilde{\sigma}_{x}^{\left(  B\right)
}\right)
\end{array}
\right]  \text{.} \label{oh}%
\end{equation}
As a final remark, we stress that the density matrices (\ref{strange1}) and
(\ref{oh}) cannot be parametrized by means of the decomposition in
(\ref{parametrization1}). For them it is needed the most general
parametrization for a two-qubit system with Hilbert space $%
\mathbb{C}
^{2}\otimes%
\mathbb{C}
^{2}$. It can be shown that any state $\hat{\rho}$ for such a system may be
parametrized as \cite{fano},%
\begin{equation}
\hat{\rho}=\frac{1}{4}\left[  I_{4\times4}+\vec{c}\cdot\vec{\sigma}\otimes
I_{2\times2}+I_{2\times2}\otimes\vec{d}\cdot\vec{\sigma}+\sum_{j\text{, }%
k=1}^{3}\gamma_{jk}\sigma_{j}\otimes\sigma_{k}\right]  \text{,}%
\end{equation}
where $\vec{c}$, $\vec{d}\in%
\mathbb{R}
^{3}$ and $\gamma_{jk}$ are \emph{real} numbers.

\section{Additional example for distinguishable particles}

In this Appendix, we consider the case of distinguishable particles with an
unbalanced scenario where the momenta of the pair are well-defined and
maximally entangled in the rest frame while the spins of the pair are
described by a one-parameter ($\xi$) family of non-maximally entangled
bipartite states. Unlike the distinguishable case considered in Section IV, we
consider a composite quantum system described in the rest frame $S$ by a total
wave-vector that differs from (\ref{as0}) and characterized by a set of
momentum constraints that differs from that in (\ref{set2}).

Consider the following set of working hypotheses on momentum states,%
\begin{equation}
\vec{p}_{A_{1}}=-\vec{p}_{A_{2}}\text{, }\vec{p}_{B_{1}}=-\vec{p}_{B_{2}%
}\text{ and, }\vec{p}_{A_{1}}=a\vec{p}_{B_{2}}\text{,} \label{set1}%
\end{equation}
where $\alpha=\frac{m_{e}}{m_{\mu}}$. This expression for $\alpha$ is obtained
by imposing that the two angles of Wigner's rotations $\varphi_{A}$ and
$\varphi_{B}$ be equal (for an explicit derivation of this specific expression
for $\alpha$, we refer to Eq. (\ref{look}) in Section V). The momentum states
$\left\vert \vec{p}_{A_{1}}\right\rangle $ and $\left\vert \vec{p}_{A_{2}%
}\right\rangle $ are the eigenvectors of $\tilde{\sigma}_{z}^{\left(
A\right)  }$ with eigenvalues $+1$ and $-1$, respectively, while $\left\vert
\vec{p}_{B_{1}}\right\rangle $ and $\left\vert \vec{p}_{B_{2}}\right\rangle $
are the eigenvectors of $\tilde{\sigma}_{z}^{\left(  B\right)  }$ with
eigenvalues $-1$ and $+1$, respectively. Thus, we consider%
\begin{equation}
\left\vert \vec{p}_{A_{1}}\right\rangle =\left\vert 0\right\rangle _{A}%
\equiv\binom{1}{0}\text{, }\left\vert \vec{p}_{A_{2}}\right\rangle =\left\vert
1\right\rangle _{A}\equiv\binom{0}{1}\text{, }\left\vert \vec{p}_{B_{1}%
}\right\rangle =\left\vert 1\right\rangle _{B}\equiv\binom{0}{1}\text{ and,
}\left\vert \vec{p}_{B_{2}}\right\rangle =\left\vert 0\right\rangle _{B}%
\equiv\binom{1}{0}\text{. }%
\end{equation}
In such case, the wave-vector $\left\vert \Psi\right\rangle _{S}$ in the rest
frame $S$ reads,%
\begin{align}
\left\vert \Psi_{e\mu}^{\left(  \text{new}\right)  }\right\rangle  &
=\left\vert \Psi\right\rangle _{S}\overset{\text{def}}{=}\frac{1}{\sqrt{2}%
}\left[  \left\vert \vec{p}_{A_{1}}\text{, }\vec{p}_{B_{1}}\right\rangle
\otimes\left\vert 0\right\rangle _{S}+\left\vert \vec{p}_{A_{2}}\text{, }%
\vec{p}_{B_{2}}\right\rangle \otimes\left\vert 0\right\rangle _{S}\right]
\nonumber\\
& \nonumber\\
&  =\frac{1}{\sqrt{2}}\left\vert 10\right\rangle \left[  \sqrt{1-\xi
}\left\vert \uparrow\downarrow\right\rangle \pm\sqrt{\xi}\left\vert
\downarrow\uparrow\right\rangle \right]  \mp\frac{1}{\sqrt{2}}\left\vert
01\right\rangle \left[  \sqrt{1-\xi}\left\vert \uparrow\downarrow\right\rangle
\pm\sqrt{\xi}\left\vert \downarrow\uparrow\right\rangle \right]  \text{.}
\label{qqq}%
\end{align}
Note that for $\xi=\frac{1}{2}$, $\left\vert \Psi\right\rangle _{S}$ in
(\ref{qqq}) becomes the wave-vector studied in \cite{suda},%
\begin{equation}
\left\vert \Psi_{\pm}\right\rangle \overset{\text{def}}{=}\frac{1}{\sqrt{2}%
}\left\vert 10\right\rangle \left\vert \phi_{\pm}\right\rangle \mp\frac
{1}{\sqrt{2}}\left\vert 01\right\rangle \left\vert \phi_{\pm}\right\rangle
\text{,} \label{q4}%
\end{equation}
with $\left\vert \phi_{\pm}\right\rangle $ defined in (\ref{69}). Focusing on
$\left\vert \Psi_{+}\right\rangle $ (the same results are obtained using
$\left\vert \Psi_{-}\right\rangle $ and following the same line of reasoning
presented in Section V, we obtain that the spin and momentum reduced density
operators in the rest frame $S$ become,%
\begin{align}
\rho_{\text{spin}}^{\left(  S\right)  }\left(  \xi\right)   &  =\frac{1}%
{4}\left[
\begin{array}
[c]{c}%
I_{2\times2}^{\left(  A\right)  }\otimes I_{2\times2}^{\left(  B\right)
}+\left(  1-2\xi\right)  \sigma_{z}^{\left(  A\right)  }\otimes I_{2\times
2}^{\left(  B\right)  }-\left(  1-2\xi\right)  I_{2\times2}^{\left(  A\right)
}\otimes\sigma_{z}^{\left(  B\right)  }+2\sqrt{\xi\left(  1-\xi\right)
}\sigma_{x}^{\left(  A\right)  }\otimes\sigma_{x}^{\left(  B\right)  }+\\
\\
+2\sqrt{\xi\left(  1-\xi\right)  }\sigma_{y}^{\left(  A\right)  }\otimes
\sigma_{y}^{\left(  B\right)  }-\sigma_{z}^{\left(  A\right)  }\otimes
\sigma_{z}^{\left(  B\right)  }%
\end{array}
\right] \nonumber\\
& \nonumber\\
&  =\left(
\begin{array}
[c]{cccc}%
0 & 0 & 0 & 0\\
0 & 1-\xi & \sqrt{\xi\left(  1-\xi\right)  } & 0\\
0 & \sqrt{\xi\left(  1-\xi\right)  } & \xi & 0\\
0 & 0 & 0 & 0
\end{array}
\right)  \text{, }%
\end{align}
and,%
\begin{equation}
\rho_{\text{momentum}}^{\left(  S\right)  }=\frac{1}{4}\left[  I_{2\times
2}^{\left(  A\right)  }\otimes I_{2\times2}^{\left(  B\right)  }-\tilde
{\sigma}_{x}^{\left(  A\right)  }\otimes\tilde{\sigma}_{x}^{\left(  B\right)
}-\tilde{\sigma}_{y}^{\left(  A\right)  }\otimes\tilde{\sigma}_{y}^{\left(
B\right)  }-\tilde{\sigma}_{z}^{\left(  A\right)  }\otimes\tilde{\sigma}%
_{z}^{\left(  B\right)  }\right]  =\frac{1}{2}\left(
\begin{array}
[c]{cccc}%
0 & 0 & 0 & 0\\
0 & 1 & -1 & 0\\
0 & -1 & 1 & 0\\
0 & 0 & 0 & 0
\end{array}
\right)  \text{,}%
\end{equation}
respectively. The concurrence of $\rho_{\text{spin}}^{\left(
\text{rest-frame}\right)  }\left(  \xi\right)  $ and $\rho_{\text{momentum}%
}^{\left(  \text{rest-frame}\right)  }$ are,%
\begin{equation}
\mathcal{C}_{\text{spin}}^{\left(  S\right)  }\left(  \xi\right)  =\sqrt
{4\xi\left(  1-\xi\right)  }\text{ and }\mathcal{C}_{\text{momentum}}^{\left(
S\right)  }=1\text{,}%
\end{equation}
respectively. The Lorentz-transformed spin density operator in (\ref{t-spin})
reads,%
\begin{align}
\rho_{\text{spin}}^{\left(  S^{\prime}\right)  }\left(  \xi\right)   &
=\frac{1}{4}\left[
\begin{array}
[c]{c}%
I_{2\times2}^{\left(  A\right)  }\otimes I_{2\times2}^{\left(  B\right)
}+\left(  1-2\xi\right)  \sigma_{z}^{\left(  A\right)  }\otimes I_{2\times
2}^{\left(  B\right)  }-\left(  1-2\xi\right)  I_{2\times2}^{\left(  A\right)
}\otimes\sigma_{z}^{\left(  B\right)  }+2\sqrt{\xi\left(  1-\xi\right)  }%
\cos2\varphi\sigma_{x}^{\left(  A\right)  }\otimes\sigma_{x}^{\left(
B\right)  }+\\
\\
+2\sqrt{\xi\left(  1-\xi\right)  }\cos2\varphi\sigma_{y}^{\left(  A\right)
}\otimes\sigma_{y}^{\left(  B\right)  }-\sigma_{z}^{\left(  A\right)  }%
\otimes\sigma_{z}^{\left(  B\right)  }%
\end{array}
\right] \nonumber\\
& \nonumber\\
&  =\left(
\begin{array}
[c]{cccc}%
0 & 0 & 0 & 0\\
0 & 1-\xi & \sqrt{\xi\left(  1-\xi\right)  }\cos2\varphi & 0\\
0 & \sqrt{\xi\left(  1-\xi\right)  }\cos2\varphi & \xi & 0\\
0 & 0 & 0 & 0
\end{array}
\right)  \text{.}%
\end{align}
Similarly, the Lorentz-transformed momentum density operator in
(\ref{t-momentum}) is given by%
\begin{align}
\rho_{\text{momentum}}^{\left(  S^{\prime}\right)  }\left(  \xi\right)   &
=\frac{1}{4}\left[
\begin{array}
[c]{c}%
I_{2\times2}^{\left(  A\right)  }\otimes I_{2\times2}^{\left(  B\right)
}-\cos2\varphi\tilde{\sigma}_{x}^{\left(  A\right)  }\otimes\tilde{\sigma}%
_{x}^{\left(  B\right)  }-\cos2\varphi\tilde{\sigma}_{y}^{\left(  A\right)
}\otimes\tilde{\sigma}_{y}^{\left(  B\right)  }+\\
\\
+\left(  1-2\xi\right)  \sin2\varphi\left(  \tilde{\sigma}_{y}^{\left(
A\right)  }\otimes\tilde{\sigma}_{x}^{\left(  B\right)  }-\tilde{\sigma}%
_{x}^{\left(  A\right)  }\otimes\tilde{\sigma}_{y}^{\left(  B\right)
}\right)  -\tilde{\sigma}_{z}^{\left(  A\right)  }\otimes\tilde{\sigma}%
_{z}^{\left(  B\right)  }%
\end{array}
\right] \nonumber\\
& \nonumber\\
&  =\frac{1}{2}\left(
\begin{array}
[c]{cccc}%
0 & 0 & 0 & 0\\
0 & 1 & -\left[  \cos2\varphi+i\left(  1-2\xi\right)  \sin2\varphi\right]  &
0\\
0 & -\left[  \cos2\varphi-i\left(  1-2\xi\right)  \sin2\varphi\right]  & 1 &
0\\
0 & 0 & 0 & 0
\end{array}
\right)  \text{.} \label{strange1}%
\end{align}
Furthermore, omitting technical details, it can be shown that the concurrences
of the Lorentz-transformed spin and momentum density matrices in the moving
reference frame $S^{\prime}$ are given by,%
\begin{equation}
\mathcal{C}_{\text{spin}}^{\left(  S^{\prime}\right)  }\left(  \xi\text{,
}\varphi\right)  =\sqrt{4\xi\left(  1-\xi\right)  }\left\vert \cos
2\varphi\right\vert \text{,} \label{1a}%
\end{equation}
and,%
\begin{equation}
\mathcal{C}_{\text{momentum}}^{\left(  S^{\prime}\right)  }\left(  \xi\text{,
}\varphi\right)  =\sqrt{1-4\xi\left(  1-\xi\right)  \sin^{2}2\varphi}\text{,}
\label{2a}%
\end{equation}
respectively.

We remark that although Eqs. (\ref{qqq}) and (\ref{set1}) differ from Eqs.
(\ref{as0}) and (\ref{set2}), respectively, we have obtained the same final
conclusions in both distinguishable cases considered: Eqs. (\ref{1a}) and
(\ref{2a}) are equal to Eqs. (\ref{ONE})\ and (\ref{TWO}), respectively. We
also point out that when $\xi=\frac{1}{2}$ and $m_{e}=m_{\mu}$ (two
electrons), we recover the main result appeared in {\cite{suda}. Finally, we
emphasize that }for distinguishable particles, the change in entanglement for
the momenta is not the same as the change in entanglement for spins (see Eqs.
(\ref{1a}) and (\ref{2a}) in addition to Figures $3$, $4$, $5$, $6$).

\end{document}